\documentclass[prd,showpacs,preprintnumbers,amsmath,amssymb,superscriptaddress,floatfix,nofootinbib]{revtex4}
\usepackage{graphicx}
\usepackage{epstopdf}
\usepackage{amsmath}
\usepackage{amsfonts}
\usepackage{amssymb}
\usepackage{bm}

\begin{document}

\title{Triangle singularity mechanism for the $pp \to \pi^+ d$ fusion reaction}

\author{Natsumi Ikeno}
\email{ikeno@tottori-u.ac.jp}
\affiliation{Department of Agricultural, Life and Environmental Sciences, Tottori University, Tottori 680-8551, Japan}
\affiliation{Departamento de F\'{\i}sica Te\'orica and IFIC,
Centro Mixto Universidad de Valencia-CSIC Institutos de Investigaci\'on de Paterna, Aptdo.22085, 46071 Valencia, Spain}

\author{Raquel Molina}
\email{raquel.molina@ific.uv.es}
\affiliation{Departamento de F\'{\i}sica Te\'orica and IFIC,
Centro Mixto Universidad de Valencia-CSIC Institutos de Investigaci\'on de Paterna, Aptdo.22085, 46071 Valencia, Spain}

\author{Eulogio Oset}
\email{oset@ific.uv.es}
\affiliation{Departamento de F\'{\i}sica Te\'orica and IFIC,
Centro Mixto Universidad de Valencia-CSIC Institutos de Investigaci\'on de Paterna, Aptdo.22085, 46071 Valencia, Spain}
\date{\today}

\begin{abstract}
We develop a model for the $pp \to \pi^+ d$ reaction based on the $pp \to \Delta(1232) N$ transition followed by $\Delta(1232) \to \pi N'$ decay and posterior fusion of $N N'$ to give the deuteron. We show that the triangle diagram depicting this process develops a triangle singularity leading to a large cross section of this reaction compared to ordinary fusion reactions. The results of the calculation also show that the process is largely dominated by the $pp$ system in $L=2, S=0$, which transfers $J=2$
to the final $\pi^+ d$ system. This feature is shown to be well suited to provide $L=2,S=1$, $J^\mathrm{tot}=3$ for $np$ in the $np(I=0) \to \pi^- pp$  followed by  $pp \to \pi^+ d$ reaction, which has been proposed recently, as a means of describing the so far assumed dibaryon $d^*(2380)$ peak. 
\end{abstract}

\maketitle
\section{Introduction}
\begin{figure}
\centering
 \includegraphics[scale=0.9]{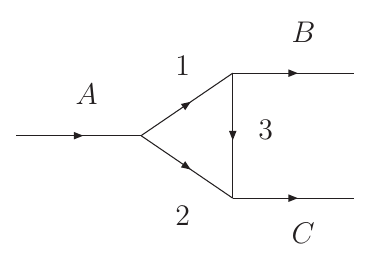}
 \caption{Feynman diagram representing the triangle singularity (TS) phenomena.}
 \label{fig:TS}
\end{figure}

Triangle singularities (TS) were introduced in Refs.~\cite{Karplus:1958zz,Landau:1959fi} and became fashionable in the 60's. In terms of Feynman diagrams, they stem from a diagram with a loop with three intermediate particles which develops a singularity when the three intermediate particles can be placed simultaneously on shell and they are collinear in a way as to satisfy the Coleman-Norton theorem~\cite{Coleman:1965xm}. This is easily stated by saying that an original particle $A$ decays into particles 1 and 2, particle 1 decays into $B$ and 3 and particles 2 and 3 merge into $C$, as shown in Fig. \ref{fig:TS}, but this process occurs in a way that all particles 1, 2, 3 are collinear in the $A$ rest frame and 3 and 2 go in the same direction, with 3 going faster than 2 and catching up to make the fusion possible. In other words, the mechanism depicted in the Feynman diagram can occur at a classical level.
A modern formulation of the problem, both intuitive and practical, is given in Ref.~\cite{Bayar:2016ftu} and the conditions for a TS to occur are condensed in a single easy equation (Eq.~(18) of Ref.~\cite{Bayar:2016ftu}).

Examples of such singularities in physical processes were searched for with no success at that time~\cite{Booth:1961zz,anisovich}, but the vast amount of experimental information collected nowadays has produced a revival of the idea identifying many present phenomena in terms of triangle singularities.

A turning point in this direction in recent times was given in the study of the $\eta(1405) \to f_0(980) \pi^0$ decay~\cite{BESIII:2012aa} solved in terms of a triangle singularity in Refs.~\cite{Wu:2011yx,Aceti:2012dj,Wu:2012pg,Achasov:2015uua}. Another example can be found in the enhancement of the $\gamma d \to K \Lambda(1405)$ cross section around $\sqrt{s} = 2110$~MeV~\cite{Moriya:2013hwg}, interpreted in terms of a triangle singularity in Ref.~\cite{Wang:2016dtb}. Also, the $\pi N(1535)$ production channel in the $\gamma p \to p  \pi^0 \eta$ reaction~\cite{Gutz:2014wit} was shown in Ref.~\cite{Debastiani:2017dlz} to be a consequence of a triangle singularity. A recent application of the TS was also the explanation of the COMPASS peak, originally associated with a new resonance, the $a_1(1420)$, in terms of a TS~\cite{Liu:2015taa,Ketzer:2015tqa,Aceti:2016yeb,compassnew}. Many other examples of TS are given in Refs.~\cite{Liang:2019jtr,Dai:2018hqb,raquel} and in a recent review on the subject in Ref.~\cite{guosakai}.

In the present work, we perform the calculations for the $pp \to \pi^+ d $ reaction which offers a very good example of a TS. There is one more reason to study this reaction: a recent work~\cite{dibaryon} proposes and explanation for the ``$d^*(2380)$'' dibaryon peak, observed in the $np \to \pi^0 \pi^0 d$, $np \to \pi^+ \pi^- d$ reactions~\cite{clement1,clement2,clement3,clementrev}, based on an old idea~\cite{barnir} that the dominant two pion production plus fusion mechanism comes from a two step single pion production process, $np \to \pi^- pp$ followed by $pp \to \pi^+ d$ (plus $np \to \pi^+ nn$ followed by $nn \to \pi^- d$ ).  

In Ref.~\cite{dibaryon}, the idea is retaken and, using recent data for the $np(I=0) \to \pi^- pp $ reaction~\cite{isoscalar,talina}, plus data for the $pp \to \pi^+ d$ one~\cite{serre}, a peak with the characteristics of the one observed in Refs~\cite{clement1,clement2,clement3} is obtained, which makes unnecessary the hypothesis of introducing a dibaryon in order to explain the $ np \to \pi^0 \pi^0 d$ ($ np \to \pi^+ \pi^- d$) peak. The large strength observed for the $np \to \pi^+ \pi^- d$ peak, of about $0.5$~mb, is made possible thanks to the size of the $pp \to \pi^+ d $ cross section, of the order of $3-4$~mb, which is abnormally large for a fusion reaction~\cite{dillig}. A novelty of this article is to show that this is tied to the existence of a TS.

The $pp \to \pi^+ d$ reaction was studied in its time reversal form, $\pi^+ d \to pp$, in Refs.~\cite{riska,green,weise} using a Quantum Mechanical formulation for the reaction using deuteron wave functions. The reaction was shown to be driven by $\Delta(1232)$ excitation and, lacking a field theoretical formulation of the problem, the TS was not identified. However, in Ref.~\cite{green} it was shown that the cross section was blowing up in the limit of the $\Delta$ width going to zero. This is a characteristic of the TS. We follow the work of Ref.~\cite{weise} closely in the dynamics used for the $\Delta$ excitation, but do a Field Theoretical formulation which allows us to identify the TS and understand why the cross section obtained is large compared with other fusion reactions. A different approach to the problem is done in \cite{schiff}, where a fully covariant formalism is developed and the different amplitudes are parameterized and fitted to the data.

The Coleman-Norton theorem expressed for the present case can be understood in the following way: the $pp$ system produces a $\Delta$ and a nucleon $N$, back to back in the $pp$ rest frame. The $\Delta$ decays into $\pi N'$, with the $\pi$ in the direction of the $\Delta$ and $N'$ in its opposite direction, which is then the direction of $N$. The $N'$ goes faster than $N$ (implicit in Eq.~(18) of Ref.~\cite{Bayar:2016ftu}) and after a while catches up with $N$ and they fuse to give the deuteron. This natural possibility, inherent to a TS, makes the cross section large, unlike other fusion reactions which rely upon large momentum components of the deuteron wave function, or equivalently, very far off shell nucleons in the intermediate states of the loop.

\section{Formalism}\label{sec:form}
\subsection{The triangle mechanism}
\begin{figure}
\centering
\includegraphics[scale=1.0]{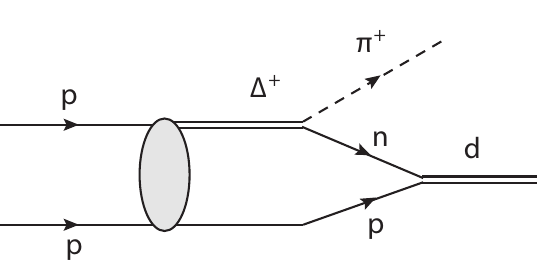}
\caption{ Mechanism for pion production with $\Delta$ excitation and $np$ fusion in the deuteron. }
 \label{fig:1}
\end{figure}

Let us study the $pp \to \pi^+ d$ reaction through $\Delta$ excitation in the intermediate state. Ignoring for the moment the dynamics of $\Delta$ excitation, the basic mechanism is depicted in Fig.~\ref{fig:1}.

The triangle diagram contains a $\Delta^+$, a neutron and a proton. The TS appears in this diagram if we can place simultaneously on shell the $\Delta^+$, the $n$, and $p$ particles, with the $\pi^+$ momentum in the direction of the $\Delta^+$ and the $n$ in the direction of the intermediate $p$ and moving faster than it, such that they can meet after some time and fuse. All those conditions are encoded in the equation (Eq.~(18) of Ref.~\cite{Bayar:2016ftu}) 
\begin{equation}
 q_{\rm on} =  q_{a_-},
\label{eq:TS_condition}
\end{equation}
where $q_{\rm on}$ is the $\Delta^+$ momentum in the $pp$ rest frame when $\Delta^+$ and the intermediate proton are placed on shell, and $q_{a_-}$ is one of the solutions when the intermediate $n$, $p$ are placed on shell matching the $d$ energy (the one where the $n$ moves faster than the $p$). Analytical formulae for $q_{\rm on}$ and $q_{a_-}$ are given in Ref.~\cite{Bayar:2016ftu}.
Technically, with the deuteron bound by 2.2~MeV this condition cannot be fulfilled, but this is no obstacle for the amplitude to develop a large strength by continuity. In practice one can see where the singularity would appear by taking the $d$ slightly unbound, and Eq.~(\ref{eq:TS_condition}) tells us that a peak, when the $\Delta$ width is negligible, should appear around $\sqrt{s} = 2179$~MeV. With the consideration of the $\Delta$ width, the singularity becomes a broad peak and experimentally this peak is seen around $\sqrt{s} = 2165$~MeV~\cite{serre}.

\subsection{Explicit model with pion exchange}
\begin{figure}
\centering
\includegraphics[scale=0.7]{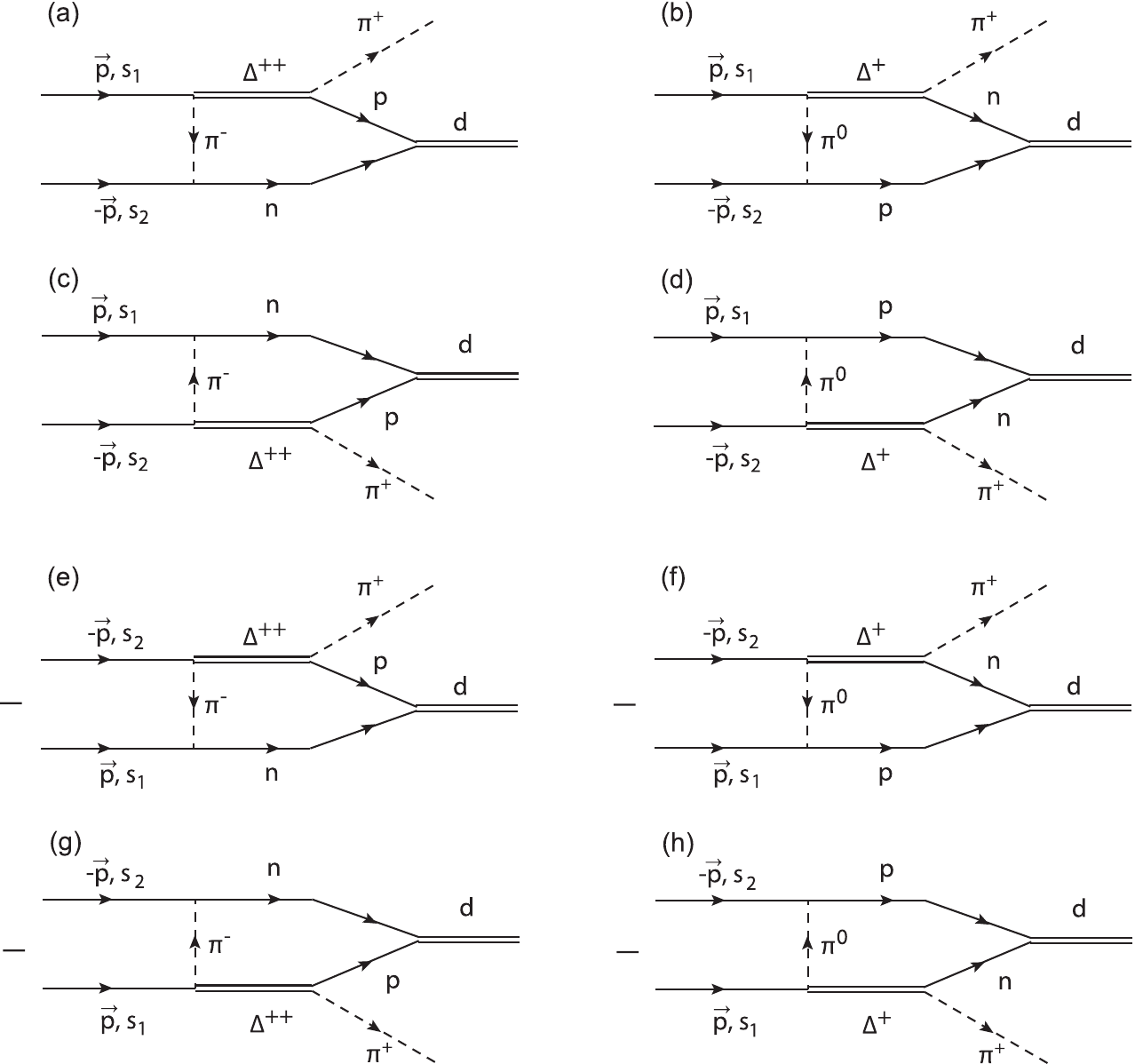}
\caption{ Diagrams contributing to $pp \to \pi^+ d$ through $\Delta$ excitation via $\pi$ exchange. }
 \label{fig:2}
\end{figure}

We start with an antisymmetrized $pp$ system with 
\begin{equation}
 |pp \rangle \equiv \frac{1}{\sqrt{2}} \left( |\vec{p},s_1 ;  -\vec{p},s_2 \rangle 
- |-\vec{p},s_2 ;  \vec{p},s_1 \rangle \right),
\label{eq:pp}
\end{equation} 
with $\vec{p} $ the momentum of one proton in the $pp$ rest frame and $s_1$, $s_2$ their spin third components. We also write the spin isospin wave function of the deuteron as
\begin{equation}
 |d \rangle \equiv \frac{1}{\sqrt{2}} | pn - np \rangle \ \chi_d, \label{eq:d}
\end{equation}
with $\chi_d$ any of the three spin 1 states ($\uparrow \uparrow$, $\frac{1}{\sqrt{2}}(\uparrow \downarrow + \downarrow \uparrow)$, $\downarrow  \downarrow$).
The Feynman diagrams that contribute to the $p p \to \pi^+ d$ process are depicted in Fig.~\ref{fig:2}.

\begin{figure}
\centering
\includegraphics[scale=0.6]{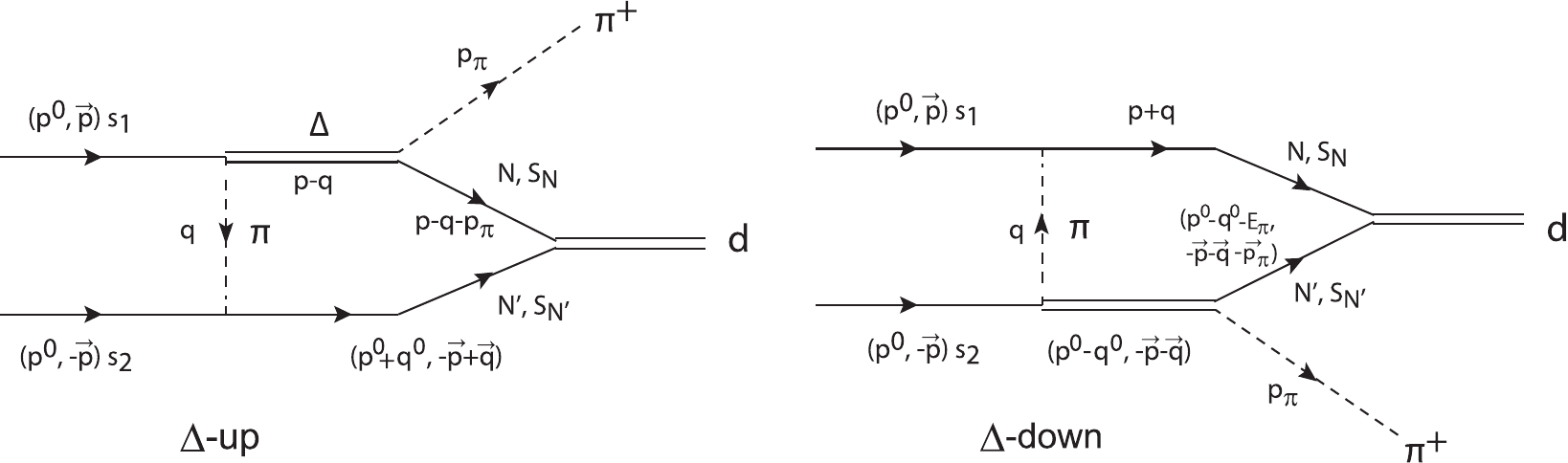}
\caption{The two topological structures obtained summing diagrams Fig.~\ref{fig:2} (a), \ref{fig:2} (b) (\ref{fig:2} (g), \ref{fig:2} (h)) in $\Delta$-up and Fig.~\ref{fig:2} (c), \ref{fig:2} (d) (and \ref{fig:2} (e), \ref{fig:2} (f)) in $\Delta$-down.}
\label{fig:3}
\end{figure}

Note, however, that diagrams (g), (h), (e) and (f) are topologically equivalent to (a), (b), (c) and (d), respectively, considering the sign of $pn$ and $np$ in the deuteron wave function, Eq.~(\ref{eq:d}) and that the spin function is symmetric.
 Hence, we can keep diagrams (a), (b), (c), (d) with a global factor $\frac{1}{\sqrt{2}} \cdot 2$. In addition, we can sum diagrams (a), (b) which have the same structure and (c), (d) taking into account the isospin coefficients. The two topologies obtained when summing (a) and (b), and (c) and (d), are depicted in Fig. \ref{fig:3} (left) and (right) respectively.

The dynamics of the process is given by the $\pi NN$ and $\pi N\Delta $ vertices. The first one is given by,
\begin{equation}
 -i \delta H_{\pi N N} = \frac{f}{m_\pi} \vec{\sigma}\cdot \vec{q} \ \tau^\lambda ; \hspace{5mm} f=1.00,
\label{eq:H_piNN}
\end{equation}
for a $\pi$ entering the $N$ line with momentum $\vec{q}$, with $\vec{\sigma}$, $\vec{\tau}$ the spin, isospin Pauli matrices and $\lambda$ the pion isospin in spherical basis. We follow the isospin convention with the pion multiplet ($-\pi^+$, $\pi^0$, $\pi^-$). Recoil corrections to Eq.~(\ref{eq:H_piNN}) are negligible and as in Ref.~\cite{weise} we do not consider them here.
The $\pi N \Delta$ vertex for $\pi N \to \Delta$ is given by
\begin{equation}
 -i \delta H_{\pi N \Delta} =  \frac{f^*}{m_\pi} \vec{S^{\dag}} \cdot \vec{q} \ T^{\dag \lambda} ; \hspace{5mm} f^* = 2.13,
\label{eq:H_piNDelta}
\end{equation}
with $\vec{S^{\dag}}$, $T^{\dag \lambda}$ the spin, isospin transition operator from spin, isospin $\frac{1}{2}$ to $\frac{3}{2}$ normalized as
\begin{equation}
\langle \frac{3}{2}, M_\Delta  | \vec{S^{\dag}_\nu} |\, \frac{1}{2}, m \rangle 
= \mathcal{C} (\frac{1}{2} \ 1 \ \frac{3}{2} ;~ m, \nu, M_\Delta)
\end{equation}
with $\nu$ the spherical index of $\vec{S^{\dag}}$, and similarly for the second $\pi N\Delta$ vertex, with the operator  $\vec{S} \cdot \vec{q} \ T_\nu$. We use the property
\begin{equation}
\sum_{M_\Delta} \langle m' | S_i | M_\Delta  \rangle   \langle  M_\Delta | S_j^{\dag} | m  \rangle 
= \langle m' | \left( \frac{2}{3}\delta_{ij} - \frac{i}{3}\epsilon_{ijk} \sigma_k \right)  | m  \rangle .
\end{equation}
\begin{table}
\caption{Matrix elements for the $\tau^{\lambda}$, $T^{\dag \lambda}$ coefficients.}\label{tab:ic}
 \begin{center}
 \setlength{\tabcolsep}{0.5em}{\renewcommand{\arraystretch}{1.7} 
  \begin{tabular}{c|c}
   \hline
   Reaction&Isospin factor ($I$)\\
   \hline
   $\pi^0 p \to p$&$1$\\
   $\pi^0 n \to n$&$ -1$\\
   $\pi^+ n \to p$&$\sqrt{2}$\\
   $\pi^- p \to n$&$\sqrt{2}$\\
   $\pi^+ p \to \Delta^{++}$&$ -1$ \\
   $\pi^+ n \to \Delta^{+}$&$ - \frac{1}{\sqrt{3}}$\\
   $\pi^0 p \to \Delta^{+}$&$\sqrt{\frac{2}{3}}$\\
   $\pi^0 n \to \Delta^{0}$&$\sqrt{\frac{2}{3}}$\\
   $\pi^- p \to \Delta^{0}$&$\frac{1}{\sqrt{3}}$\\
   $\pi^- n \to \Delta^{-}$&$1$\\
   \hline
  \end{tabular}}
 \end{center}
\end{table}

To ease the calculations we provide the matrix elements of the $\tau^{\lambda}$, $T^{\dag \lambda}$ coefficients in Table \ref{tab:ic}.

 Taking into account the matrix elements of $\tau^\lambda$, $T^\lambda$, the factor $\frac{2}{\sqrt{2}}$ from the weights of the diagrams of Fig.~\ref{fig:2} and the isospin sign of the $pn$, $np$ components of the deuteron in Eq.~(\ref{eq:pp}) (the spin is symmetric and does not change by exchange $pn \to np$), we get the weights $ h_{\Delta-{\rm up}} = \frac{4 \sqrt{2}}{3}$, $ h_{\Delta-{\rm down}} = -\frac{4 \sqrt{2}}{3}$ for the two topological structures,  $\Delta$-up and $\Delta$-down of Fig. \ref{fig:3}.

We then find
\begin{eqnarray}
-it_{\Delta-{\rm up}} =&& \frac{4}{3} \sqrt{2} \left( \frac{f^*}{m_\pi} \right)^2  \left( \frac{f}{m_\pi} \right) 
\int \frac{d^4q}{(2\pi)^4} (-)\vec{S}_1 \cdot \vec{p}_\pi  (-) \vec{S}^{\dag}_1 \cdot \vec{q} \
\vec{\sigma}_2 \cdot \vec{q} \ (-i) \ g_d \
\theta(q_{\rm max} - |\vec{p}\,^{\rm CM}_d|) \nonumber\\
&\times &  \frac{2M_\Delta}{2 E_\Delta (\vec{p} - \vec{q}\,)} \ \frac{i}{p^0-q^0- E_\Delta (\vec{p} - \vec{q}\,) +i\frac{ \Gamma_\Delta}{2}} \ \frac{i}{q^{02} - \vec{q}~^2 - m^2_\pi + i\epsilon}
\nonumber\\
&\times &  \frac{2M_N}{2 E_N (\vec{p} - \vec{q} - \vec{p}_{\pi})} 
\ \frac{i}{p^0 - q^0  - E_\pi(\vec{p}_{\pi}) -  E_N (\vec{p} - \vec{q} - \vec{p}_{\pi} ) + i \epsilon} 
\nonumber\\
&\times &
\frac{2M_N}{2 E_N(-\vec{p} + \vec{q}\,)} 
\ \frac{i}{p^0 + q^0 -  E_N(- \vec{p} + \vec{q}\,) + i \epsilon} ,
\end{eqnarray}
where the subindices 1, 2 in the spin operators refer to the upper and lower baryon in the diagram $\Delta$-up of Fig.~\ref{fig:3}. $E_N (\vec{p}_N) = \sqrt{M^2_N + \vec{p}\,^2_N}$, $E_\pi (\vec{p}_\pi) = \sqrt{m^2_\pi + \vec{p}\,^2_\pi} $. In field theory, the deuteron appears as a coupling $g_d$ and a $\theta(q_{\rm max} - |\vec{p}\,^{\rm CM}_d|)$ where $\vec{p}\,^{\rm CM}_d$ is the nucleon momentum of the deuteron in the $d$ rest frame (see details in Appendix~\ref{App:A} )
 
Similarly, we can write
\begin{eqnarray}
-it_{\Delta-{\rm down}} =&-& \frac{4}{3} \sqrt{2} \left( \frac{f^*}{m_\pi} \right)^2  \left( \frac{f}{m_\pi} \right) 
\int \frac{d^4q}{(2\pi)^4} \vec{\sigma}_1 \cdot \vec{q} \ (-)\vec{S}_2 \cdot \vec{p}_\pi  (-)\vec{S}^{\dag}_2 \cdot \vec{q} 
\ (-i) \ g_d \
\theta(q_{\rm max} - |\vec{p'}\,^{ \rm CM}_d|) \nonumber\\
&\times &  \frac{2M_\Delta}{2 E_\Delta (-\vec{p} - \vec{q}\,)} \ \frac{i}{p^0-q^0- E_\Delta (-\vec{p} - \vec{q}\,) +i\frac{ \Gamma_\Delta}{2}} \ \frac{i}{q^{02} - \vec{q}~^2 - m^2_\pi + i\epsilon}
\nonumber\\
&\times &  \frac{2M_N}{2 E_N(-\vec{p} - \vec{q} - \vec{p}_{\pi})} 
\ \frac{i}{p^0 - q^0 - E_\pi(\vec{p}_{\pi}) - E_N(-\vec{p} - \vec{q} - \vec{p}_{\pi} ) + i \epsilon}\nonumber\\
&\times&
\ \frac{2M_N}{2 E_N(\vec{p} + \vec{q}\,)} 
\ \frac{i}{p^0 + q^0 -  E_N( \vec{p} + \vec{q}\,) + i \epsilon}.
\end{eqnarray}

The CM nucleon momenta of the deuteron in the $\Delta$-up and $\Delta$-down mechanisms is obtained with $\frac{1}{2}$ times the difference of the nucleon momenta incoming in the deuteron
\begin{eqnarray}
 \vec{p}\,^{\rm CM}_d  &=& \vec{p} - \vec{q} - \frac{\vec{p}_{\pi}}{2},\label{eq:cm1} \\
 \vec{p'}_d ^{\rm CM} &=& - \vec{p} - \vec{q} - \frac{\vec{p}_{\pi}}{2}. 
 \label{eq:cm2}
\end{eqnarray}

In order to establish a link with the wave function formalism of Refs.~\cite{riska,green,weise}, it is convenient to write 
\begin{eqnarray}
& & \frac{1}{p^0 - q^0  - E_\pi -  E_N (\vec{p} - \vec{q} - \vec{p}_{\pi} ) + i \epsilon} 
\ \frac{1}{p^0 + q^0 -  E_N(- \vec{p} + \vec{q}\,) + i \epsilon}  \nonumber\\
=&& \frac{1}{2p^0 - E_\pi -  E_N (- \vec{p} + \vec{q} )  - E_N (\vec{p} - \vec{q} - \vec{p}_{\pi} ) + i \epsilon}
\nonumber\\
&\times& \left\{ 
\frac{1}{p^0 - q^0 - E_\pi -  E_N (\vec{p} - \vec{q} - \vec{p}_{\pi} ) + i \epsilon} 
+ \frac{1}{p^0 + q^0 -   E_N (- \vec{p} + \vec{q} )  + i \epsilon} 
\right\},
\end{eqnarray}

\begin{equation}
 \frac{1}{q^{2} - m^2_\pi } = 
 \frac{1}{2\omega} \left\{  \frac{1}{q^0 - \omega(q)  + i\epsilon}
- \frac{1}{q^0 + \omega(q)  - i\epsilon} \right\} ; \hspace{5mm} 
\omega(q) = \sqrt{\vec{q}\,^2 + m^2_\pi },
\end{equation}

This reduces in two the number of factors $(q^0 - \alpha)^{-1}$ in the different terms and allows an immediate $q^0$ integration using Cauchy's residues. We get
\begin{eqnarray}
-it_{\Delta-{\rm up}} = - g_d
\frac{4 \sqrt{2}}{3}  \left( \frac{f^*}{m_\pi} \right)^2  \left( \frac{f}{m_\pi} \right) 
\int \frac{d^3q}{(2\pi)^3} \vec{S}_1 \cdot \vec{p}_\pi \  \vec{S}^{\dag}_1 \cdot \vec{q} \
\vec{\sigma}_2 \cdot \vec{q} \  F(\vec{p}, \vec{q}, \vec{p}_\pi) {\cal F}_\pi(\vec{q}\,)
\label{eq:t_Delta_up}
\end{eqnarray}
where we have added a form factor ${\cal F}_\pi(\vec{q}\,)=\left(\frac{\Lambda_\pi^2 - m_\pi^2}{\Lambda_\pi^2 + \vec{q}\,^2}\right)^2 $, with values of $\Lambda_\pi$ around $1-1.2$ GeV, and 
$F(\vec{p}, \vec{q}, \vec{p}_\pi)$ is given in Appendix~\ref{App:newD}.
 With the choice of momenta in $\Delta$-down of Fig. \ref{fig:3}, the integral over the set of propagators is easily done, simply changing $\vec{p} \to -\vec{p} $, and we obtain for the sum of the two terms,
\begin{eqnarray}
-it^{\pi} =- g_d
\frac{4 \sqrt{2}}{3}  \left( \frac{f^*}{m_\pi} \right)^2  \left( \frac{f}{m_\pi} \right) 
\int \frac{d^3q}{(2\pi)^3} {\cal F}_\pi(\vec{q}\,)
\left\{
\vec{S}_1\cdot \vec{p}_\pi \  \vec{S}_1^\dagger  \cdot \vec{q} \
\vec{\sigma}_2 \cdot \vec{q} \ F(\vec{p}, \vec{q}, \vec{p}_\pi) \ 
-
\vec{\sigma}_1 \cdot \vec{q} \ \vec{S}_2 \cdot \vec{p}_\pi \  \vec{S}^{\dag}_2 \cdot \vec{q} 
 \ F(-\vec{p}, \vec{q}, \vec{p}_\pi) \ 
\right\}\nonumber\\
\label{eq:t_Delta}
\end{eqnarray}

It should be noted that we have considered the full pion propagator (integrating over its $q^0$ variable), and have not done the usual static approximation used in most works $( q^{02} - \vec{q}\,^2 - m^2_\pi )^{-1} \to (- \vec{q}\,^2 - m^2_\pi )^{-1}$ including Refs.~\cite{riska,green,weise}.
The matrix elements of the spin operators are calculated in Appendix~\ref{App:B}, as $Q_{ij}^{\rm up}$, $Q_{ij}^{\rm down}$ for $i =  \uparrow \uparrow,  \uparrow \downarrow$, and $j =  \uparrow \uparrow,$ $\frac{1}{\sqrt{2}} ( \uparrow \downarrow +  \downarrow \uparrow ),$ $\downarrow \downarrow$, and we can write
\begin{eqnarray}
-it^{\pi}_{ij} = -g_d 
\frac{4 \sqrt{2}}{3}  \left( \frac{f^*}{m_\pi} \right)^2  \left( \frac{f}{m_\pi} \right) 
\int \frac{d^3q}{(2\pi)^3}  {\cal F}_\pi(\vec{q}\,)
 \left\{ 
Q_{ij}^{\rm (up)} F(\vec{p}, \vec{q}, \vec{p}_\pi) -
Q_{ij}^{\rm (down)} F(-\vec{p}, \vec{q}, \vec{p}_\pi)
\right\}.
\end{eqnarray}
The momenta $\vec{p}_\pi$, $\vec{q}$ in the spin operators are boosted to the $\Delta$ rest frame giving $\vec{p}\,'_\pi$, $\vec{q}\,'$, as shown in Appendix \ref{App:C}. 
One also needs to multiply by two the sum and average over spins of $|t^\pi|^2$  to account for the initial states $\downarrow \downarrow$, $\downarrow \uparrow$ contributions. Thus,
\begin{equation}
 \overline \sum \sum |t^\pi|^2 = 2 \frac{1}{4} \sum_{i,j} |t^{\pi}_{ij}|^2 = \frac{1}{2} \sum_{i,j} |t^{\pi}_{ij}|^2 ,\label{eq:sumt}
\end{equation}
and the cross section for $pp \to \pi^+ d$ is then given by
\begin{equation}
 \frac{d \sigma}{d \cos \theta_\pi} =  \frac{1}{4\pi} \frac{1}{s} (M_N)^2 \ M_d \ \frac{p_\pi}{p} \  \overline{\sum} \sum  |t^\pi|^2, \label{eq:cs}
\end{equation}
where $\cos \theta_\pi$ is $\displaystyle \frac{\vec{p} \cdot \vec{p}_\pi}{ |\vec{p}\,| |\vec{p}_\pi|}$
(see Eq.~(\ref{eq:AppB_def_moment}) for the expressions of $\vec{p}$, $\vec{p}_\pi$ and $\vec{q}$\,). 

Now it is easy to establish the connection with the ordinary deuteron wave function \cite{machleidt}. Using Eq.~(\ref{eq:wf}) of Appendix~\ref{App:A} (with the normalization of Eq.~(\ref{eq:Normwf}) ), considering that we have $\int d^3p/(2\pi)^3$ integrations instead of $\int d^3p$ in Eq.~(\ref{eq:Normwf}), and including the weight factors of Field Theory ($M/E$ which are very close to unity) we have
\begin{eqnarray}
& & \frac{M_N}{ E_N(-\vec{p} + \vec{q}\,)} \ \frac{M_N}{ E_N (\vec{p} - \vec{q} - \vec{p}_{\pi})} 
\ \frac{1}{2p^0 - E_\pi -  E_N (- \vec{p} + \vec{q} )  - E_N (\vec{p} - \vec{q} - \vec{p}_{\pi} ) + i \epsilon}
\ g_d
\
\theta(q_{\rm max} - |\vec{p} - \vec{q} - \frac{\vec{p}_{\pi}}{2} |) \nonumber\\
&\to& (-) (2 \pi)^{3/2} \ \psi (|\vec{p} - \vec{q} - \frac{\vec{p}_{\pi}}{2} |)\ .\label{eq:pro}
\end{eqnarray}
The sign is needed since the expression of Eq.~(\ref{eq:wfd}) is negative and the wave function of Ref.~\cite{machleidt} is positive.

In the expression for $F(\vec{p}, \vec{q}, \vec{p}_\pi)$ in Eq.~(\ref{eq:F_pqpi}), the $\Delta$ width appears, and we consider it energy dependent. Details are given in Appendix~\ref{App:D}.

\subsection{Effect of short range correlations, $g'$ term}
As done in Ref.~\cite{osetweise} when we have terms like
\begin{equation}
 S^{\dag}_{i} q_i \sigma_j q_j  \ \frac{1}{q^2 - m^2_\pi}\label{eq:sdag}
\end{equation}
approximately,
\begin{equation}
 q_i q_j \frac{1}{q^2 - m^2_\pi} \simeq q_i q_j \frac{1}{-\vec{q}\,^2 } 
= (q_i q_j - \frac{1}{3} \vec{q}\,^2  \delta_{ij} ) \frac{1}{- \vec{q}\,^2}
+ \frac{1}{3} \frac{\vec{q}\,^2}{-\vec{q}\,^2} \delta_{ij}.\label{eq:apro}
\end{equation}
The last term is a $\delta$ function in coordinate space which becomes inoperative when there is a strong short range repulsion between the baryons. The $NN$ and $N \Delta$ wave functions have a correlation factor at short distances vanishing at $r \to 0$ that kills the $\delta$
function. To kill this term it suffices to add a term $\frac{1}{3} \delta_{ij}$ to $q_i q_j/(q^2 - m^2_\pi)$. In the realistic case, taking into account nuclear correlation functions for the evaluation, one finds it is necessary to add the term~\cite{osetweise}
\begin{equation}
 g' \delta_{ij},\label{eq:corr}
\end{equation}
with values of $g' \simeq 0.6$, which comes from the modifications to $\pi$ and $\rho$ exchange (studied in the next subsection), together with the form factors from these correlations. To evaluate this term, we replace
\begin{equation}
\vec{S} \cdot \vec{p}_{\pi} \  \vec{S}^{\dag} \cdot \vec{q} \
\vec{\sigma} \cdot \vec{q} \   \frac{1}{q^2 - m^2_\pi}
\to g' \ \vec{S} \cdot \vec{p}_{\pi} \ \vec{S}^{\dag} \cdot \vec{\sigma}.
\end{equation} 

Once again we evalulate the spin matrix elements in Appendix~\ref{App:B} and the $d^4 q$ integral is simpler than before since we do not have the pion propagator and we find:
\begin{eqnarray}
-it^{\rm corr}_{ij} = -g_d 
\frac{4 \sqrt{2}}{3}  \left( \frac{f^*}{m_\pi} \right)^2  \left( \frac{f}{m_\pi} \right) 
\ g' \int \frac{d^3q}{(2\pi)^3}  {\cal F}_\pi(\vec{q}\,)
 \left\{ 
Q_{ij}^{\prime \rm (up)} F'(\vec{p}, \vec{q}, \vec{p}_\pi) -
Q_{ij}^{\prime  \rm (down)} F^\prime (-\vec{p}, \vec{q}, \vec{p}_\pi)
\right\},
\label{eq:t_corr}
\end{eqnarray}
with 
$F'(\vec{p}, \vec{q}, \vec{p}_\pi)$ a function which we write in Appendix~\ref{App:newD}
and $Q_{ij}^{\prime \rm (up)}=Q_{ij}^{\prime \rm (down)}$ as shown in Appendix~\ref{App:B}.

\subsection{ $\rho$-exchange}
We can use the $\pi$-exchange formalism with the substitution
\begin{eqnarray}
\frac{f^*}{m_\pi} \ \frac{f}{m_\pi} 
\ \vec{S}^{\dag}_1 \cdot \vec{q} \  \vec{\sigma}_2 \cdot \vec{q}\,{\cal F}_\pi(\vec{q}\,) \ 
\frac{1}{q^2 - m^2_\pi + i \epsilon}
\to 
\frac{f_\rho^*}{m_\rho} \ \frac{f_\rho}{m_\rho} 
(\vec{S}^{\dag}_1 \times \vec{q}) \
(\vec{\sigma}_2 \times \vec{q})\,{\cal F}_\rho(\vec{q}\,) \ 
\frac{1}{q^2 - m^2_\rho + i \epsilon}\ ,\label{eq:piex}
\end{eqnarray}
with ${\cal F}_\rho(\vec{q}\,)=\left(\frac{\Lambda_\rho^2 - m_\rho^2}{\Lambda_\rho^2 + \vec{q}\,^2}\right)^2 $. 
As in Ref.~\cite{weise} we take the values 
\begin{equation}
f_\rho = 7.96;~~~ f^*_\rho = 13.53,\label{eq:fval}
\end{equation}
and play with $\Lambda_\pi$ and $\Lambda_\rho$ parameters with $\Lambda_\pi\sim 1.1~\mathrm{GeV};~~~ \Lambda_\rho\sim 1.8~\mathrm{GeV}$ used in Ref~\cite{weise}.

Hence for  $\Delta$-up we have the spin operator
\begin{eqnarray}
 \vec{S}_1 \cdot \vec{p}\,'_{\pi} \ (\vec{S}^{\dag}_1 \times \vec{q}\,') \cdot
(\vec{\sigma}_2 \times \vec{q}\,) 
&=&  S_{1,i} \, p'_{\pi, i} \, \epsilon_{jkm} \, S^\dagger_{1,k} \, q'_{m} \, \epsilon_{j \ell n}
\, \sigma_{2, \ell} \, q_n \nonumber\\
&=& (\delta_{k \ell} \, \delta_{m n} - \delta_{k n} \, \delta_{\ell m} )\,
p'_{\pi, i} \left( \frac{2}{3} \delta_{ik} - \frac{i}{3} \epsilon_{iks} \sigma_{1,s}
\right) q'_m q_n \, \sigma_{2, \ell}
\nonumber\\
&=& 
\left( \frac{2}{3} \delta_{ik} - \frac{i}{3} \epsilon_{iks} \sigma_{1,s}
\right) p'_{\pi, i}  \,
( \vec{q}\,' \cdot \vec{q}\, \sigma_{2,k} - q_{k} \, \vec{\sigma}_{2} \cdot \vec{q}\,' )\,
\nonumber\\
&=& 
\frac{2}{3} \vec{q}\,' \cdot \vec{q} \  \vec{p}\,'_{\pi} \cdot  \vec{\sigma}_2
- \frac{2}{3} \,  \vec{p}\,'_{\pi}  \cdot \vec{q} \  \vec{\sigma}_{2} \cdot \vec{q}\,'
- \frac{i}{3} \vec{q}\,' \cdot \vec{q} \, \epsilon_{iks}\,  p'_{\pi, i} \sigma_{2,k} \sigma_{1,s}
+\frac{i}{3} \epsilon_{iks}\,  
p'_{\pi, i} q_k\sigma_{1,s} \, \vec{\sigma}_{2} \cdot \vec{q}\,'  ,
\nonumber\\
\label{eq:dups}
\end{eqnarray}
where we already write the expressions in term of the boosted $\vec{p}\,'_{\pi}$ and 
$\vec{q}\,'$ momenta (see Appendix \ref{App:C}). The two operators with $\epsilon_{iks}$ in the last line of Eq.~(\ref{eq:dups}) are evaluated in Appendix~\ref{App:B}, the first one in Eq. (\ref{eq:twosig}) and the second one in Eq.~(\ref{eq:onesig}).
The results for the matrix elements $Q^{(\rho, {\rm up})}_{ij}$,  $Q^{(\rho, {\rm down})}_{ij}$ are shown in Appendix~\ref{App:B}. Then we write
\begin{eqnarray}
-it^{\rho}_{ij} =- g_d \frac{4 \sqrt{2}}{3}  
\left( \frac{f^*}{m_\pi} \right)  \left( \frac{f^*_\rho}{m_\rho} \right) 
 \left( \frac{f_\rho}{m_\rho} \right) 
\int \frac{d^3q}{(2\pi)^3}  {\cal F}_\rho(\vec{q}\,)
 \left\{ 
Q_{ij}^{(\rho, {\rm up)}} F(\vec{p}, \vec{q}, \vec{p}_\pi, \rho) -
Q_{ij}^{(\rho, {\rm down)}} F(-\vec{p}, \vec{q}, \vec{p}_\pi, \rho)
\right\},
\end{eqnarray}
where $F(\vec{p}, \vec{q}, \vec{p}_\pi, \rho)$ has the same expression as Eq.~(\ref{eq:F_pqpi}) simply changing $m_\pi \to m_\rho$.

\subsection{Impulse approximation}
\begin{figure}
\centering
\includegraphics[scale=0.6]{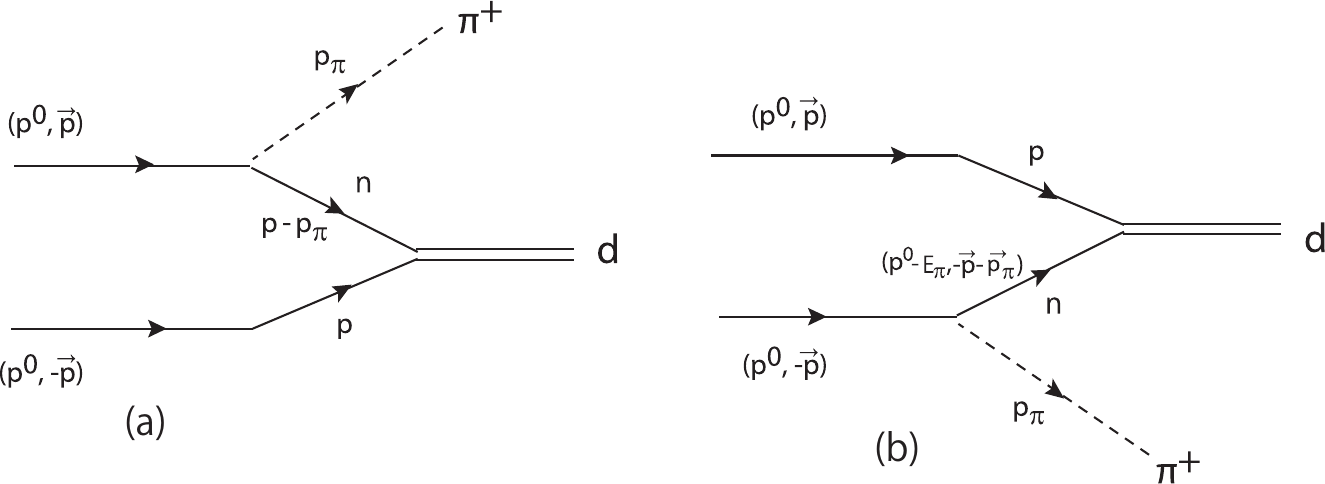}
\caption{Diagrams for the impulse approximation. }
\label{fig:4}
\end{figure}

As in Refs.~\cite{riska,green,weise}, we also consider the impulse approximation contribution corresponding to the diagrams of Fig.~\ref{fig:4}.
The $\pi^+$ emission vertex is given by
\begin{equation}
 - i\delta H_{\pi^+np} = - \sqrt{2}\frac{f}{m_\pi} \vec{\sigma} \cdot \vec{p}_\pi,\label{eq:tpi}
\end{equation}
and in our formalism we have for the pion up (a) and pion down (b) diagrams.
\begin{eqnarray}
&&-it_{I}^{\rm up} = \frac{\sqrt{2}f}{m_\pi} \vec{\sigma}_1 \cdot \vec{p}_\pi \ 
 \frac{M_N}{ E_N(\vec{p} - \vec{p}_\pi)} \ \frac{1}{ p^0 - E_\pi  - E_N (\vec{p} -  \vec{p}_{\pi} ) + i \epsilon}  \ g_d \ \theta(q_{\rm max} - |\vec{p} -\frac{\vec{p}_{\pi}}{2} |),\label{eq:diagup}
\\
&&-it_{I}^{\rm down} = -\frac{\sqrt{2}f}{m_\pi} \vec{\sigma}_2 \cdot \vec{p}_\pi \ 
 \frac{M_N}{ E_N(-\vec{p} - \vec{p}_\pi)} \ \frac{1}{ p^0 - E_\pi  - E_N (-\vec{p} -  \vec{p}_{\pi} ) + i \epsilon} \ g_d \ \theta(q_{\rm max} - |-\vec{p} -\frac{\vec{p}_{\pi}}{2} |),\label{eq:diagdo}
\end{eqnarray}
where $(\vec{p} -\frac{\vec{p}_{\pi}}{2} )$ or $(-\vec{p} -\frac{\vec{p}_{\pi}}{2} )$ stand for the momenta of the deuteron in the CM frame for the up and down mechanisms, respectively. The factor $\sqrt{2}$ in Eq. (\ref{eq:tpi}) is cancelled by the $1/\sqrt{2}$ factor of the deuteron function in Eq.~(\ref{eq:d}), and we have added in Eqs.~(\ref{eq:diagup}), (\ref{eq:diagdo}), the factor $\sqrt{2}$ that appears when antisymmetizing the $pp$ system in Eq.~(\ref{eq:pp}) and summing the corresponding diagrams coming from using the antisymmetrized pair of protons as given in Eq.~(\ref{eq:pp}).
However, there is a problem here because $|\vec{p} -\frac{\vec{p}_{\pi}}{2} |$, or $|-\vec{p} -\frac{\vec{p}_{\pi}}{2} |$ are always bigger than $q_{\rm max}$. While for the triangle mechanisms discussed before, small momenta of the deuteron (see Eqs.~(\ref{eq:cm1}) and~(\ref{eq:cm2})) are allowed and the formalism works well (we shall compare with results using explicitly the Bonn deuteron wave function \cite{machleidt}), here we are forced to use explicitly the deuteron wave function at large momenta, which unavoidably contains uncertainties, but we shall see that the effects of the IA are very small. It is easy to rewrite the former equations in term of the deuteron wave functions. For this, we recall Eq.~(\ref{eq:wf}) from Appendix~\ref{App:A} and write (note that $E_N(p) = p^0$), similarly to Eq.~(\ref{eq:pro}),
\begin{equation}
  g_d   \, \frac{M_N}{ E_N(\vec{p} \,)}  \, \frac{M_N}{ E_N(\vec{p} - \vec{p}_{\pi}\,)} 
 \ \frac{ \theta(q_{\rm max} - |\vec{p} -\frac{\vec{p}_{\pi}}{2} |)}{ 2p^0-E_\pi - E_N(\vec{p})  - E_N (\vec{p} -  \vec{p}_{\pi} ) + i \epsilon}
= - \sqrt{(2 \pi)^3} \ \psi (\vec{p} -\frac{\vec{p}_{\pi}}{2})\ .
\label{eq:wfd}
\end{equation}
Then, defining the factor for the impulse approximation 
\begin{eqnarray}
 F_I=\frac{\sqrt{2}f}{m_\pi}\frac{E(\vec{p})}{M_N}\sqrt{(2\pi)^3}
 \label{eq:fi}
\end{eqnarray}
we obtain  expressions, $-i t^I_{ij}$, for the spin transitions including the up and down diagrams of Fig.~\ref{fig:3} , which we write explicitly in Appendix~\ref{App:E}.
Altogether the final cross section is given by Eq. (\ref{eq:cs}) substituting $|t^\pi|$ by $|t|^2$. Thus, in Eq.~(\ref{eq:sumt}) we make the replacement
\begin{eqnarray}
 \sum_{ij} |t_{ij}|^2\longrightarrow\sum_{ij}|t^{\pi}_{ij}+t^{\rho}_{ij}+t^\mathrm{corr.}_{ij}+t^I_{ij}|^2\label{eq:t2}
\end{eqnarray}
We thus sum $t^\pi$, $t^\rho$, $t^\mathrm{corr.}$, $t^I$ coherently in the amplitudes. We shall study the effect of each one of the terms and also will separate the contribution of the different spin transitions.

\section{Results}

\begin{figure}
\centering
\includegraphics[scale=0.9]{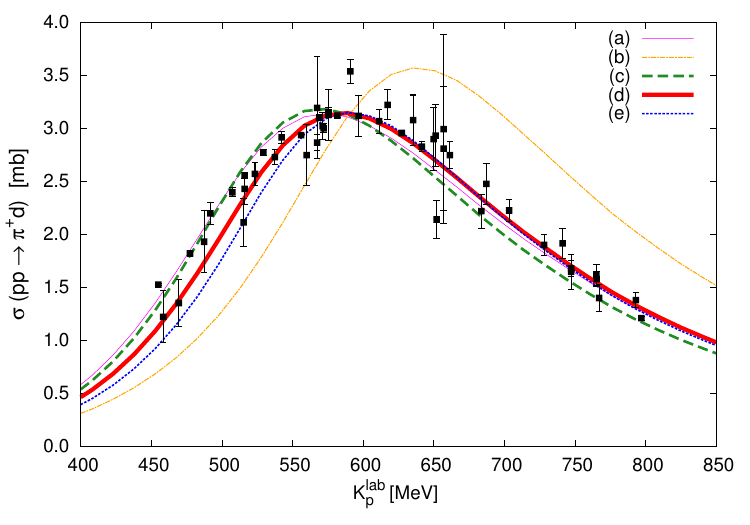}
\caption{Cross section of $pp\to\pi^+d$ as a function of the kinetic energy in the lab frame of the proton. The variable $s$ is $s=4M_N^2+2M_N K_p^\mathrm{lab}$.}
\label{fig:fig5}
\end{figure}

 In this section we show the results from fits to the data of the $pp \to \pi^+ d$ reaction with the model described in the previous section. There are data from Ref.~\cite{serre} on the $pp \to \pi^+ d$ reaction, but we also take into account equivalent data for the time reversal reaction $\pi^+ d \to pp $. Actually,
the two sets of data are available in the SAID data base of Ref.~\cite{said} in terms of $\pi^+ d \to pp$ cross sections. We convert these data into $ pp \to \pi^+ d$ cross sections using the detailed balance theorem
\begin{equation}
 \sigma_{pp \to \pi^+ d } = 2 \ \frac{3}{4} \left( \frac{p_\pi}{p} \right)^2 \sigma_{\pi^+ d \to pp},
\label{eq:new}
\end{equation}
where $p_\pi$, $p$, appearing in the expression of the $pp \to \pi^+ d$ cross section of Eq.~(\ref{eq:cs}), are the pion and proton momenta in the $pp$ rest frame. The factor 2 in Eq.~(\ref{eq:new}) accounts for the factor $\frac{1}{2}$ due to the identity of the protons in the $\pi d \to pp$ reaction  and the factor $\frac{3}{4}$ accounts for the number of spin polarization of the $d$ and the $pp$ systems. The data converted into $pp \to \pi^+ d$ cross sections are plotted in Fig.~\ref{fig:fig5} and come from Refs.~\cite{Aebischer:1976gs,Rogers:1957zz,Stadler:1954zz,Ritchie:1983nt,Boswell:1982xm,Guelmez:1993nh,Shimizu:1982dx,Hoftiezer:1982sy,Alteholz:1994ze,Cohn:1957zz,Albrow:1971yy,Mathie:1983ee,Norem:1971nu,Mayer:1985cv,Nann:1979yg}.

In Fig. \ref{fig:fig5} we show the results for $\sigma(pp\to\pi^+d)$ as a function of $K_p^\mathrm{lab}$, the proton kinetic energy in the $pp$ lab frame, the variable used in the data of Ref. \cite{serre}, which are obtained by integrating $d\sigma/d\mathrm{cos}\theta_\pi$ of Eq.~(\ref{eq:cs}) over the pion angle, with the sum of all contributions given by Eq. (\ref{eq:t2}).

We perform different fits to the data that require a detailed discussion. The parameters $M_\Delta$, $\Gamma_\Delta$, $\Lambda_\pi$, $\Lambda_\rho$, are varied keeping the coupling constants fixed. Five different fits are done to the data, which are described in Table~\ref{table:fit}, and are called (a)-(e).

The first fit, fit~(a), is done to all the data in the range of $K_p^\mathrm{lab} \in$ [450:800]~MeV. This fit looks good to the sight, however, it has a reduced $\chi^2$/dof of the order of 9, which indicates a bad fit. This is the first consideration to be made. 
The $\chi^2$/dof is large because there are many data with tiny errors and some data are incompatible with other ones. Yet, what we observe is that the solution of the fit prefers a mass lower than the nominal one, $M_\Delta=1232$~MeV, and a width much larger than the experimental one of $\Gamma_\Delta=117$~MeV. 
One might be tempted to think that this experiment provides evidence of a larger $\Delta$ width than the standard one. We rather think that in the lower energy region of the data there must be other mechanisms additional to those considered by us which are responsible for the results obtained in the fit. We are not concerned about that since our purpose is to explain the data at a reasonable level to show the dominance of the $\Delta$ excitation and how the triangle singularity is exhibited. We also value much the information that we obtain concerning the different spin transitions which is new in this paper.

We might then wonder whether it is possible to obtain a reasonable fit to the data fixing the mass and width of the $\Delta$ to the nominal ones. This is done in fit~(b). The results are absolutely unacceptable as one can see in Fig.~\ref{fig:fig5}, with a reduced $\chi^2$/dof of the order of 330. As one can see in Fig.~\ref{fig:fig5}, the data is definitely demanding a smaller $\Delta$ mass, closer to the pole mass of the PDG~\cite{pdg}, as if the triangle singularity selected this mass rather than the Breit Wigner mass.

In view of this, we conduct another fit to the data  restricting the range to reasonable values of the $\Delta$ mass and width, $M_\Delta \in$ [1200:1250] and $\Gamma_\Delta \in$ [100:150], and this is fit~(c), which does not differ much from fit (a) at low proton energies but reduces the cross section a bit above the $\Delta$ peak.

Having admitted that we should not push our model to be too accurate with the data at low proton energies, we make two new fits, restricting the range of the data to $K_p^{\mathrm lab} \in$ [525:700] and [550:700]~MeV, which select the data closer to the $\Delta$ peak. These fits are called (d) and (e). As we can see, both fits give a similar mass of around 1215~MeV and the width of fit~(d) is $\Gamma_\Delta = 150$~MeV while for fit~(e) is $\Gamma_\Delta = 117$~MeV. The $\chi^2$/dof values have improved considerably when removing points from the region where our model would require other contributions. While fit~(e) to the restricted data is acceptable, when observed in Fig.~\ref{fig:fig5} versus all data, it is showing a large discrepancy with some of the low energy data with small errors. If we look at fits (c)(d)(e), they provide a band that we could consider as uncertainty of our model in a fit to the data. This band region includes most of the data. In order to continue with the results for other observables given by our model, we select the fit (d) which is in the middle of the band and discuss the contribution of the different ingredients considered in the model, and then evaluate angular distributions and the cross section for the different spin transitions. Other choices within that band change very little the final results and certainly do not change at all the conclusions.

\begin{table}
 \begin{center}
\caption{ Results of the different fits done to obtain $M_\Delta$, $\Gamma_\Delta$, $\Lambda_\pi$, $\Lambda_\rho$ as free parameters except the fit (b).}
\label{table:fit}
 \setlength{\tabcolsep}{0.5em}{\renewcommand{\arraystretch}{1.7} 
  \begin{tabular}{c|c|c|c|c|c}
   \hline 
    & Range of $K_p^{\mathrm lab}$ and restrictions [MeV] & Parameters [MeV] &  N of data & $\chi^2$ & $\chi^2$/dof\\
   \hline 
   (a) & [450:800]      & $M_\Delta = 1216$    & 50 & 415.29 & 9.03\\
     & $M_\Delta \in$ [1150:1240] & $\Gamma_\Delta = 200$   & & &  \\
     & $\Gamma_\Delta \in$ [50:200] & $\Lambda_\pi = 1175$  & & &  \\
     & $\Lambda_\pi \in$ [800:1300] & $\Lambda_\rho = 1400$ & & &  \\
     &  $\Lambda_\rho \in$ [1400:1900]  &  & & &  \\
    \hline
  (b) & [450:800]     & $M_\Delta = 1232$    & 50 & 15883.25 &  330.90 \\
     & $M_\Delta$, $\Gamma_\Delta$ fixed  & $\Gamma_\Delta = 117$   & & &  \\
     &  $\Lambda_\pi \in$ [900:1200]      & $\Lambda_\pi = 1152$  & & &  \\
     &   $\Lambda_\rho \in$ [1400:1900]   & $\Lambda_\rho = 1900$ & & &  \\
    \hline
  (c) & [450:800]     & $M_\Delta = 1208$    & 50 & 792.50 &  17.23 \\
     & $M_\Delta \in$ [1200:1250]    & $\Gamma_\Delta = 150$   & & &  \\
     & $\Gamma_\Delta \in$ [100:150] & $\Lambda_\pi = 1014$  & & &  \\
     &  $\Lambda_\pi \in$ [900:1200] & $\Lambda_\rho = 1400$ & & &  \\
     &  $\Lambda_\rho \in$ [1400:1900]  &  & & &  \\
    \hline
  (d) & [525:700]      & $M_\Delta = 1215$    & 29 & 101.15 & 4.05 \\
     & $M_\Delta \in$ [1200:1250]    & $\Gamma_\Delta = 150$   & & &  \\
     & $\Gamma_\Delta \in$ [100:150] & $\Lambda_\pi = 1015$  & & &  \\
     &  $\Lambda_\pi \in$ [900:1200] & $\Lambda_\rho = 1428$ & & &  \\
     &  $\Lambda_\rho \in$ [1400:1900]  &  & & &  \\
    \hline
  (e) & [550:700]      & $M_\Delta = 1213$    & 25 & 43.10 & 2.05 \\
     & $M_\Delta \in$ [1200:1250]    & $\Gamma_\Delta = 117$   & & &  \\
     & $\Gamma_\Delta \in$ [100:150] & $\Lambda_\pi = 966$  & & &  \\
     &  $\Lambda_\pi \in$ [900:1200] & $\Lambda_\rho = 1552$ & & &  \\
     &  $\Lambda_\rho \in$ [1400:1900]  &  & & &  \\
    \hline
  \end{tabular}}
 \end{center}
\end{table}

\begin{figure}
\centering
\includegraphics[scale=0.8]{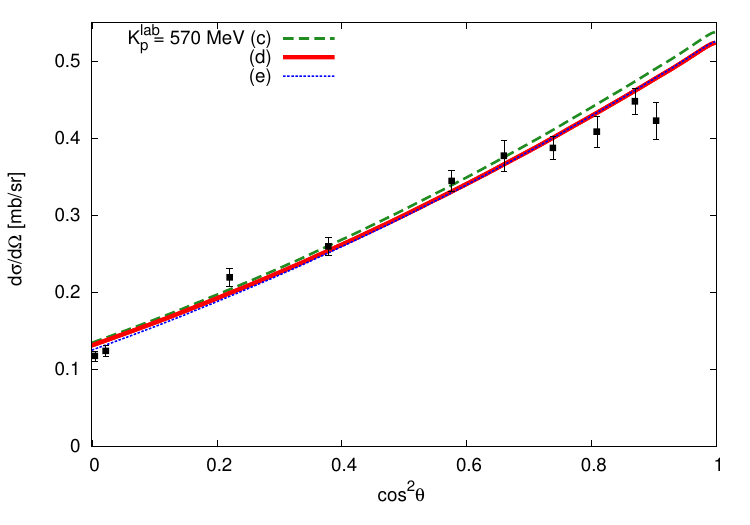}
\caption{The differential cross section, $d\sigma/d\Omega=\frac{1}{2\pi}\frac{d\sigma}{d\mathrm{cos}\theta_\pi}$, as a function of $\mathrm{cos}^2\theta_\pi$ for $K^\mathrm{lab}_p=570$ MeV, with the sets of parameters (c), (d), (e) of Table~\ref{table:fit}.}
\label{fig:fig6}
\end{figure}

\begin{figure}
\centering
\includegraphics[scale=0.9]{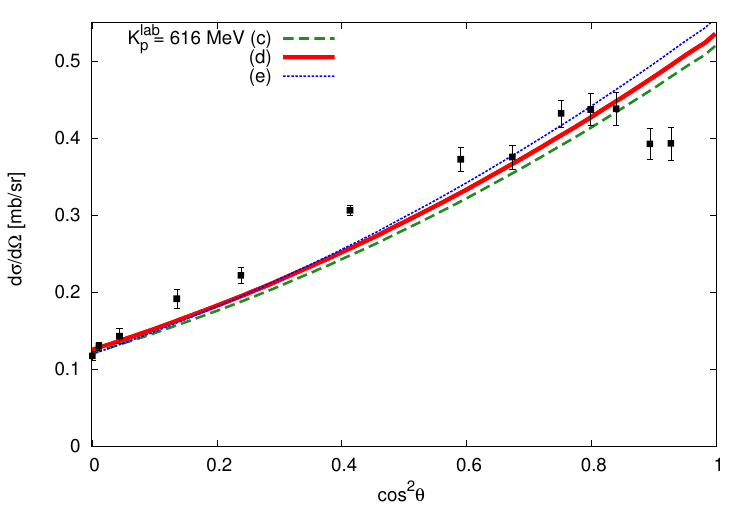}
\caption{The same as Fig. \ref{fig:fig6} for $K_p^\mathrm{lab}=616$ MeV.}
\label{fig:fig7}
\end{figure}

\begin{figure}
\centering
\includegraphics[scale=0.9]{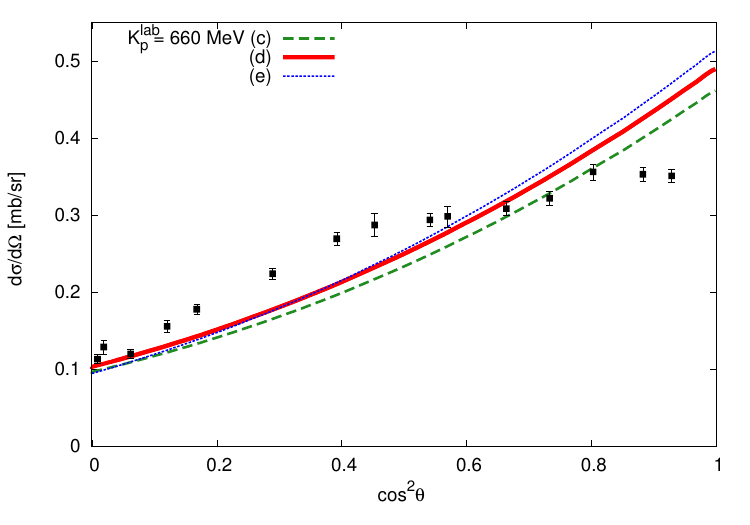}
\caption{Same as Fig. \ref{fig:fig6} for $K_p^\mathrm{lab}=660$ MeV.}
\label{fig:fig8}
\end{figure}

It is interesting to see what happens with the angular distributions. The nature of the reaction, with the two initial protons antisymmetrical, guarantees that $d\sigma/d\mathrm{cos}\theta$ will be the same for $\vec{p}_\pi$ or $-\vec{p}_\pi$, which means that it depends on $\mathrm{cos}^2\theta_\pi$ and we plot it in Figs. \ref{fig:fig6}, \ref{fig:fig7},  \ref{fig:fig8}, for different values of $K_p^\mathrm{lab}=570$, $616$ and $660$ MeV, respectively. Here we  also show the results with the three sets of parameters of the accepted error band, from the fits (c), (d), (e) of Table~\ref{table:fit}.

We can see that the global agreement is quite good, with a perfect agreement for $K_p^\mathrm{lab}=570$ MeV, and  some discrepancies at very forward angles for $K_p^\mathrm{lab}=616$ MeV and $K_p^\mathrm{lab}=660$ MeV. It is interesting to mention that the shape of our results at forward angles is very similar to the one obtained in Ref.~\cite{schiff}, where a parameterization of the amplitudes was done using a fully covariant formalism.

\begin{figure}
\centering
\includegraphics[scale=0.9]{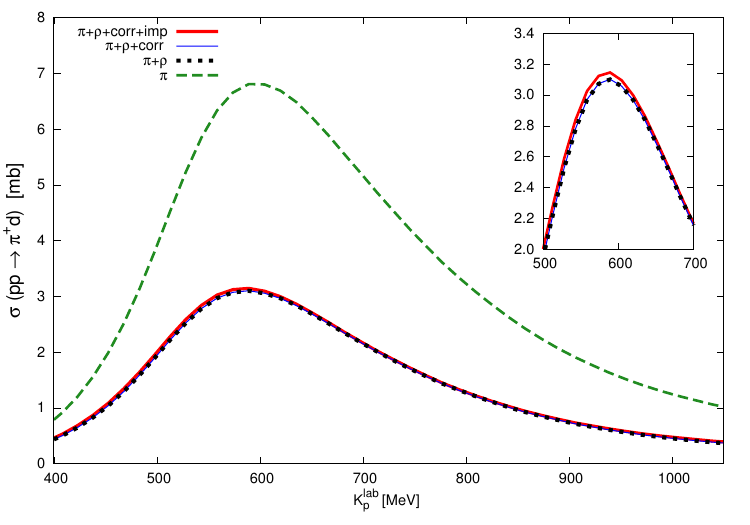}
\caption{Contribution of the different terms: The $\pi$, $\pi+\rho$, $\pi+\rho+\mathrm{corr}$ and  $\pi+\rho+\mathrm{corr}+\mathrm{imp}$. Inset: all terms except $\pi$-exchange to see the small differences. Results obtained with set (d) of Table~\ref{table:fit}.}
\label{fig:fig9}
\end{figure}

It is also interesting to see the effect of the different contributions that we have evaluated, which we show in Fig. \ref{fig:fig9}. 
What we see is that the pion exchange is the dominant term and the inclusion of the $\rho$ exchange reduces substantially the cross section, as already found in \cite{weise}, although not in \cite{green} where the $\rho$ contribution is moderate. On the other hand the effect of short range correlations, $g'$ term of Eq. (\ref{eq:corr}), is negligible and so is the effect of the impulse approximation, Eqs. (\ref{eq:diagup}), (\ref{eq:diagdo}), as we could expect when compared with the $\Delta$ triangular mechanism that gives rise to a TS. On this point we diverge from Ref.~\cite{weise} where the contribution of the impulse approximation is found individually small but sizeable when added coherently to the other terms. This, however, does not seem to be the case in Ref.~\cite{green}. We should note that in \cite{green,weise} results for the deuteron wave function of the Reid soft core potential~\cite{reid}  were used, while we rely upon the more recent Bonn potential \cite{machleidt}. Since both Refs. \cite{green,weise} investigated the time reversal reaction $\pi^+d\to pp$, the authors did not calculate the angular distributions that are presented here, which give extra support to our model. 

\begin{figure}
\centering
\includegraphics[scale=0.9]{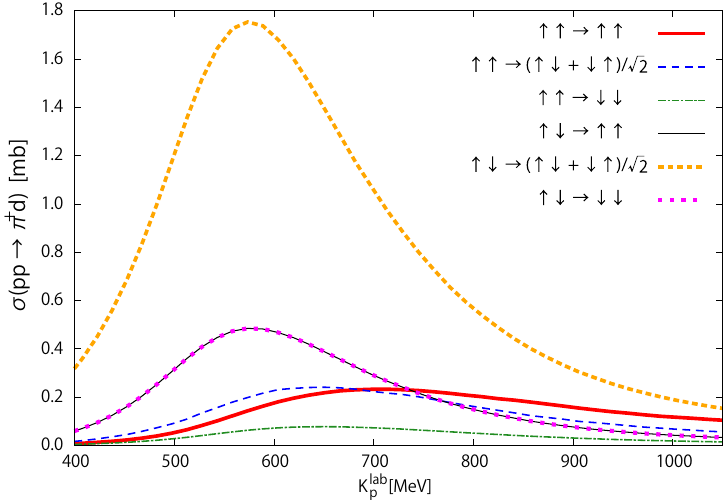}
\caption{Contribution of the different spin transitions. A factor two is included to account for transitions from $\downarrow\uparrow$ and $\downarrow\downarrow$ which are identical to those of $\uparrow\downarrow$ and $\uparrow\uparrow$ respectively. Results obtained with set (d) of Table~\ref{table:fit}.}
\label{fig:fig10}
\end{figure}

It is also very interesting to show the different contributions of the spin transitions, which are shown in Fig.~\ref{fig:fig10}.

We observe that the shape of the cross section depends on the channel. The transitions from the initial state $\uparrow\uparrow$ peak at higher energy, particularly the $\uparrow\uparrow\to \uparrow\uparrow$. However, those coming from the inital $\uparrow\downarrow$ (or $\downarrow\uparrow$ which are the same) peak around $K_p^\mathrm{lab}\simeq 600$ MeV, which is where one finds the peak of the total cross section. It is worth mentioning that the $\uparrow\downarrow\to\uparrow\uparrow$ and $\uparrow\downarrow\to\downarrow\downarrow$ transitions give the same contribution, while the largest one comes from $\uparrow \downarrow\to\frac{1}{\sqrt{2}}(\uparrow\downarrow+\downarrow\uparrow$). Altogether we can claim that most of the cross section comes from the initial $pp$ state $\uparrow\downarrow$ (or $\downarrow\uparrow$). This might be an indication that the $S=0$ contribution is dominant and indeed this is the case. Using the matrix elements of the Appendix~\ref{App:B} for the $\downarrow\uparrow$ transitions to the final deuteron spin states, we find that it is the initial state combination $\frac{1}{\sqrt{2}}(\uparrow\downarrow -\downarrow\uparrow)$ the one that is responsible for the transitions in this case. Thus, our model produces dominance of $S=0$ in the $pp$ initial state. This implies, because of the antisymmetry of the protons, that the orbital angular momentum of the protons must be even. We could have $L=0, L=2$ in our model with pion exchange with the $q_iq_j$ structure in the two vertices of the pion exchange, according to the separation in Eq.~(\ref{eq:apro}). We may wonder which one of them dominates, but this is clear because we found that the $g'$ term, which selects the $L=0$ part, gives a negligible contribution. This leaves $S=0,L=2$ as the dominant contribution for the process. One can see that for the correlation term, the transition from $\uparrow\downarrow\to\frac{1}{\sqrt{2}}(\uparrow\downarrow+\downarrow\uparrow)$, which is dominant for $\pi+\rho$ exchange, is exactly zero (the $Q'_{22}$ term). The $L$ even solution also agrees with the positive parity of the final state with $d(+)$, $\pi(-)$, and a $p$-wave coupling of the pion to the $\Delta$, up to the boost corrections that we have done. This initial state with $L=2$, $S=0$ ($^1D_2$) gives $J=2$, which means that the $\pi^+d$ system also has $J=2$, and is in the $^3P_2$ configuration, $(^{2S+1}L_J)$. It is interesting to note that this transition, $pp(^1D_2)\to \pi^+d(^3P_2)$, is also the one that was found dominant in Ref. \cite{schiff}, and also in the experimental analysis of partially polarized data in \cite{Albrow:1971yy}, and more recently in \cite{igor1,igor2}. It is also interesting to remark that, as in \cite{igor1,igor2}, the shape of the $^1D_2$ transition is very similar to the total cross section unlike for other partial waves.

At this point we would like to mention that if we make the substitution of Eq.~(\ref{eq:pro}) in the contribution of $\pi+\rho+$ correlations, the results that we obtain are remarkably similar to those obtained before. We think that such a good agreement lies in our choice of $q_\mathrm{max}$  such as to give the exact $np$ triplet scattering length, as discussed in Appendix~\ref{App:A}.

\section{Conclusions and discussion}
We have performed a calculation for the $p p \to \pi^+ d$ reaction using a model in which the two proton system  goes to $\Delta(1232) N$, the $\Delta(1232)$ decays to $\pi^+ N'$ and the two nucleons $N, N'$ fuse to produce the deuteron. We follow a Feynman diagrammatic approach in which there is a loop with three baryons in the intermediate state. We show that the mechanism develops a triangle singularity when the three baryons $\Delta, N, N'$ are placed on shell in the loop, their momenta are collinear in the $pp$ rest frame and the $N'$ nucleon travels in the same direction as $N$ but faster, such that after some time it catches up with $N$ and fuses to give the deuteron (Coleman Norton theorem). This shows that the reaction is peculiar in the sense that the deuteron is made easily, not forcing large nucleon momenta in the wave function, which happens in most fusion reactions.  This is the reason why the cross section for this reaction is very large in comparison with typical fusion reactions. The dynamics for $\Delta N $ excitation is done by means of pion and $\rho$ exchange, using standard values for the couplings and form factors. We find a good agreement with the data for the integrated cross section as a function of the $p p$ energy and the slope of the $\pi^+$ angular distribution is also well described. Some small deviation from the data for $\mathrm{cos}^2\theta_\pi$ close to 1 is observed, which was also observed in another model \cite{schiff}. 
\begin{figure}
 \centering
 \includegraphics[scale=0.75]{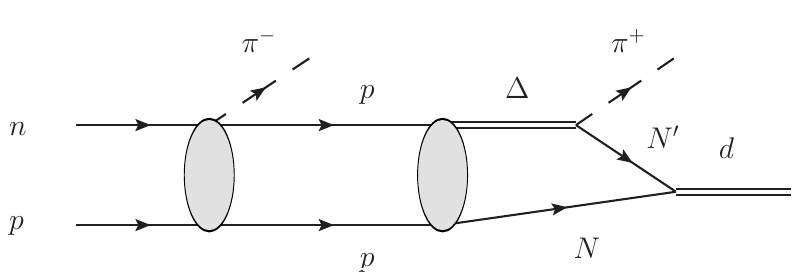}
 \caption{Two step mechanism for $np \to \pi^+\pi^-d$ investigated in \cite{dibaryon} (where $\pi^+nn$ production in the first step is also considered).}
 \label{fig:twos}
\end{figure}
    
  One novel information provided in the work is the contribution of the different spin transitions, which has allowed us to identify the most important channel for the reaction, which is $pp$ in $L=2, S=0$ (hence $J=2$), going to $\pi^+$ and $d$ with $S=1,S_z=0$. This was also observed in the experimental analysis of $pp\to\pi^+d$ using a polarized target in Refs. \cite{Albrow:1971yy,igor1,igor2}. This means that the final $\pi^+ d$ system, with $S=1$ for the deuteron and $L=1$ from a pion coupling in $p$-wave, is mostly formed with total angular momentum $J=2$. This configuration is interesting in order to interpret results related to the ``dibaryon'' peak of the $d^*(2380)$, which in a recent work has been proposed as coming from a sequential pion production mechanism, $np (I=0) \to \pi^- pp$ followed by $pp \to \pi^+ d$, see Fig.~\ref{fig:twos}. Since in the dominant term we have $L=2$ for $pp$ in the second step and the pion in the first step mostly couples in $p$-wave, the $pp$ parity is transferred to the $np$ initial state, which will have $L=even$. Consequently, since we have $np(I=0)$ the spin will be $S=1$. With similar arguments as done here, one can have dominance of $L=2$ in the first step if the transition is driven by pion exchange, and then we have $L=2, S=1$, which can couple to $J^\mathrm{tot}=1,2,3$ for the $np(I=0)$ initial state. Also having $J=2$ for the final $\pi^+ d$, together with the $\pi^-$ in $p$-wave in the first step, can equally produce the $J^\mathrm{tot}=1,2,3$ state. We should note that $L=2,S=1$ and $J^\mathrm{tot}=3$ are the favorite quantum numbers so far associated to the ``$d^*(2380)$'' dibaryon, and, thus, our picture provides a natural path to get these quantum numbers in the $np(I=0)$ original state. 
  
  So far we can have $J^\mathrm{tot}=1,2,3$ for the initial state but we can go one step further to justify the $J^{\mathrm{tot}}=3$ dominance. The key of the argumentation resides in the fact that we found that the dominant transition was from $\frac{1}{\sqrt{2}}(\uparrow\downarrow-\downarrow\uparrow)$ for $pp$ to $\frac{1}{\sqrt{2}}(\uparrow\downarrow+\downarrow\uparrow)$ for the deuteron. Hence, the deuteron is mostly formed with $S_z=0$. On the other hand, the $\pi^+$ going mostly forward or backward, as we also found, has $L_z=0$. Then we have $J=2, J_z=0$ for the $\pi^+ d$ final state. Note that we can reach the same conclusion from the $S=0$, $L=2$ $pp$ configuration, since the protons in the $z$ direction have $L_z=0$, hence the $pp$ state has $J=2$, $J_z=0$. To complete the total spin $J^{\mathrm{tot}}$ one needs now the angular momentum of the $\pi^-$ produced in the $np(I=0)\to\pi^-pp$ step. One can see that $N^*(1440)N$ and the $NN$ production in $np(I=0)\to\pi^-pp$ prior to the $\pi^-$ emission play a relevant role, which makes the pion to couple in $L=1$ in this case. Then, it is easy to see that with $N^*(1440)$ or $N$ excitation driven by pion exchange, the pions going forward for $N^*$-up ($N$-up) or backward for $N^*$-down ($N$-down), in the nomenclature given for the $\Delta$ excitation before, are preferred since this makes the pion propagator bigger. Recall that, for instance, for $N^*$-up, the $\pi$ exchanged has momentum $\vec{q}=\vec{p}\,'-\vec{p}_{N^*}$, with $\vec{p}\,'$ the momentum of the neutron in Fig.~\ref{fig:twos}, which minimizes $\vec{q}\,^2$ for $\vec{p}_{N^*}$ in the direction of $\vec{p}\,'$, and hence also $\vec{q}$ goes in this direction. Then, terms like $\vec{p}_\pi\cdot \vec{q}$ which come from the spin operators will be magnified when $\vec{p}_\pi$ and $\vec{q}$ go in the same direction. Note that the angular distributions measured in Ref.~\cite{isoscalar} show some preference for pions in the forward direction. Since $\vec{q}$ is favored in the $\vec{p}\,'$ direction, then $\vec{p}$, the momentum of the lower intermediate proton in Fig.~\ref{fig:twos} will be $-\vec{p}\,'+\vec{q}$ that goes in the $\vec{p}\,'$ direction and this also leaves the $z$ direction unchanged in the $np$ or $pp$ systems, hence, $L_z=0$ for this pion. The $|2,0\rangle$ state for $\pi^+d$ and $|1,0\rangle$ for the $\pi^-$ combine to $|J^\mathrm{tot},0\rangle$ with the Clebsch-Gordan coefficient $\mathcal{C}(2\,1\,J^\mathrm{tot};0\,0\,0)$ and we see that $J^\mathrm{tot}=2$ is forbidden and $J^\mathrm{tot}=3$ is favored. The same argumentation can be applied for the initial $S=1$, $L=2$, $np(I=0)$ pair. Since we had $\pi^+d$ in $J=2$, $J_z=0$, and the $\pi^-$ produced in $L=1,L_z=0$ we have $J_z^\mathrm{tot}=0$ for the initial state. Since $L_z=0$ for the $np$ system moving in the $z$ direction, then $S_z=0$ for $np$ and we have again the $\mathcal{C}(2\,1\,J^\mathrm{tot};0\,0\,0)$ Clebsch-Gordan coefficient, although the $2$ and $1$ now refer to $L$ and $S$. The Clebsch-Gordan coefficients squared give a factor $3/2$ from $J^\mathrm{tot}=3$ to $J^\mathrm{tot}=1$. Altogether, we arrive to the most favored production mode for the initial $np$ system: $I=0$, $S=1$, $L=2$, $J^\mathrm{tot}=3$, the $^3D_3$ partial wave where a signal of the ``$d^*(2380)$'' is seen, and the configuration $I(J^P)=0(3^+)$ common to the initial and final state in the observed peak of the $np(I=0)\to \pi^+\pi^-d$ reaction \cite{clementrev}.

     In summary, we accomplish two goals here: first we show a clear experimental example of a triangle singularity, so far not identified in previous works on the subject, and second, the dynamics of the reaction, together with the $np(I=0) \to \pi^- pp$ reaction provides a two step process that, according to the work of \cite{dibaryon}, gives a natural explanation of the peak, position and strength of the $np(I=0) \to \pi^- \pi^+ d$ reaction. The calculations done here give further information concerning spin, parity and angular momenta of the initial $np(I=0)$ state in the energy region of the peak, in agreement with experimental findings.

\appendix
\section{The deuteron in Field Theory}\label{App:A}
We follow an approach for the $NN$ scattering in $s$-wave in $J=1$, like the one a followed in Ref.~\cite{ danijuan} that can be linked to the one used in the chiral unitary approach~\cite{npa,angelsK}.
We write a potential in momentum space as
\begin{equation}
 V(q, q') = V \ \theta(q_{\rm max} - | \vec{q} |)
\ \theta(q_{\rm max} - |\vec{q}\,' |),
\label{eq:AppA_deut_V}
\end{equation}
from where the $t$ matrix satisfying the Lippmann Schwinger equation results at
\begin{equation}
 t(q, q') = t \ \theta(q_{\rm max} - | \vec{q} |)
\ \theta(q_{\rm max} - |\vec{q}\,' |), 
\end{equation}
with
\begin{equation}
t(E) = \frac{1}{V^{-1} - G(E)} ,
\label{eq:t_E}
\end{equation}
where $V$ is supposed to be energy independent and $G(E)$ is the loop function
\begin{equation}
 G(E) = \int_{|\vec{q}\,| < q_{\rm max} } d^3q \frac{1}{E - E_N(q)- E_N(q) + i\epsilon}.
\end{equation}
We take an average nucleon mass and  $M_d = 2 M_N - B$, $B =2.22$~MeV.
Since $t(E)$ has a pole at the deuteron mass, $M_d$, we can write
\begin{equation}
 t(E) = \frac{1}{G(M_d) - G(E)}.
\label{t_deut}
\end{equation}
Writing close to the pole
\begin{equation}
 t \simeq \frac{\tilde g_d^2}{E - M_d},
\end{equation}
we find that
\begin{eqnarray}
\tilde  g^2_d = \lim_{E \to M_d} (E - M_d) \ t
= \lim_{E \to M_d} \frac{E - M_d}{G(M_d) - G(E)} = \frac{1}{-\frac{\partial G(E)}{\partial E}}
\end{eqnarray}
where in the last step we have used l'H\^opital's rule.
This is the formulation in this framework of the Weinberg compositeness conditions~\cite{weinberg,baru}. The $G$ function is regularized with a cut off, $q_{\rm max}$, for $|\vec{q}\,|$ and then the $d$ wave function in momentum space is obtained as~\cite{danijuan}
\begin{eqnarray}
 \psi(\vec{p}\,)\equiv\langle \vec{p}\,|\psi\rangle = \tilde g_d \ \frac{ \theta(q_{\rm max} - | \vec{p}|) } {
E_d - E_N(p) -E_N(p) },
\label{eq:wf}
\end{eqnarray}
with a normalization 
\begin{equation}
 \int d^3p\, |\langle \vec{p}\,|\psi\rangle |^2=1.
\label{eq:Normwf}
\end{equation}
To determine $q_{\rm max}$, we follow a strategy which is to determine the scattering length from Eq.~(\ref{eq:t_E}). Using again Ref.~\cite{danijuan}, we find
\begin{equation}
 a = 2 \pi^2 M_N \, \frac{1}{G(M_d) - G(2M_N)}.
\end{equation} 
We get the experimental value $a = 5.377$~fm for $q_{\rm max} = 240$~MeV and then 
$\tilde g^2_d  = (2.68 \times 10^{-3})^2$~MeV$^{-1}$.
However, the formalism that we use in the reaction discussed in the paper relies on loop functions that contain $\displaystyle \frac{d^3q}{(2\pi)^3}$ hence in our formalism,
\begin{equation}
 g_d = (2\pi)^{3/2} \tilde g_d = (2\pi)^{3/2} \ (2.68 \times 10^{-3})~{\rm MeV}^{-1/2}.
\label{eq:q_para}
\end{equation}
It is interesting to compare this results with the standard formula of Weinberg adapted to Ref.~\cite{danijuan} normalization (see also Ref.~\cite{miguel})
\begin{equation}
 g^2_d = (2\pi)^3 \tilde g^2_d = \frac{8\pi \gamma}{M_N^2}; \hspace{5mm}
\gamma  = \sqrt{M_N B}.
\label{eq:g2_para}
\end{equation}
where $B$ is the deuteron binding energy. This formula gives $g_d = (2\pi)^{3/2} \ 2.30$~MeV$^{-1/2}$ very close to Eq.~(\ref{eq:q_para}). It is also interesting to compare the results of Eq.~(\ref{eq:wf}) for the deuteron wave function with the one of the Bonn potential~\cite{machleidt} 
\begin{equation}
 \langle \vec{p}\,|\psi\rangle = \frac{1}{N} \sum_{j} \frac{C_j}{\vec{p}\,^2 + m^2_j}
\end{equation}
with 
\begin{equation}
 N^2 = \int d^3p \left(  \sum_{j} \frac{C_j}{\vec{p}\,^2 + m^2_j} \right)^2.
\end{equation}
where the coefficients $C_j$, $m_j$, are given in Ref. \cite{machleidt}.
The agreement is remarkable up to $q_{\rm max}$ where the wave function has fallen down in about two orders of magnitude.

\section{Functions $F(\vec{p}, \vec{q}, \vec{p}_\pi)$ and $F'(\vec{p}, \vec{q}, \vec{p}_\pi) $}\label{App:newD}
The function $F(\vec{p}, \vec{q}, \vec{p}_\pi)$ entering Eq.~(\ref{eq:t_Delta_up}) is given by
\begin{eqnarray}
F(\vec{p}, \vec{q}, \vec{p}_\pi)
&=& \frac{M_N}{ E_N(-\vec{p} + \vec{q}\,)} \ \frac{M_N}{ E_N (\vec{p} - \vec{q} - \vec{p}_{\pi})} 
\ \frac{M_\Delta}{ E_\Delta (\vec{p} - \vec{q}\,)} \ \frac{1}{ 2\omega(q)}
\ \frac{1}{2p^0 - E_\pi -  E_N (- \vec{p} + \vec{q} )  - E_N (\vec{p} - \vec{q} - \vec{p}_{\pi} h) + i \epsilon}
\nonumber\\
&\cdot& \Biggl\{ \,\,
\frac{1}{p^0 - \omega(q) - E_\Delta ( \vec{p} - \vec{q}\,) +i\frac{ \Gamma_\Delta}{2}}
\ \frac{1}{p^0 - \omega(q) - E_\pi -  E_N (\vec{p} - \vec{q} - \vec{p}_{\pi} ) + i \epsilon} 
\nonumber\\
& & +
\frac{1}{p^0 - \omega(q) - E_\Delta ( \vec{p} - \vec{q}\,) +i\frac{ \Gamma_\Delta}{2}}
\ \frac{1}{2p^0 - E_\Delta( \vec{p} - \vec{q}\,) -  E_N( -\vec{p} + \vec{q}\,) + i \frac{ \Gamma_\Delta}{2}}
\nonumber\\
& & +
\frac{1}{p^0 - \omega(q) - E_N(-\vec{p} + \vec{q}\,) +i \epsilon }
\ \frac{1}{2p^0 - E_\Delta( \vec{p} - \vec{q}\,) -  E_N( -\vec{p} + \vec{q}\,) + i \frac{ \Gamma_\Delta}{2}}\,\,
\Biggr\}
\theta(q_{\rm max} - |\vec{p} - \vec{q} - \frac{\vec{p}_{\pi}}{2} |)\ , \nonumber\\
\label{eq:F_pqpi}
\end{eqnarray}
where $2p^0=\sqrt{s}$.

The function $F'(\vec{p}, \vec{q}, \vec{p}_\pi)$ entering Eq.~(\ref{eq:t_corr}) is given by
\begin{eqnarray}
F'(\vec{p}, \vec{q}, \vec{p}_\pi)
&=& \frac{M_N}{ E_N(-\vec{p} + \vec{q}\,)} \ \frac{M_N}{ E_N (\vec{p} - \vec{q} - \vec{p}_{\pi})} 
\ \frac{M_\Delta}{ E_\Delta (\vec{p} - \vec{q}\,)} 
\ \frac{1}{2p^0 - E_\pi -  E_N (- \vec{p} + \vec{q} )  - E_N (\vec{p} - \vec{q} - \vec{p}_{\pi} ) + i \epsilon}
\nonumber\\
&\cdot&  
\frac{1}{2 p^0 - E_\Delta ( \vec{p} - \vec{q}\,) - E_N( -\vec{p} + \vec{q}\,)  +i\frac{ \Gamma_\Delta}{2}}
\cdot
\theta(q_{\rm max} - |\vec{p} - \vec{q} - \frac{\vec{p}_{\pi}}{2} |).
\end{eqnarray}

\section{Spin matrix elements}\label{App:B}
We shall study explicitly the spin transitions for initial $pp$ spins $s_1$, $s_2$ to final spins of $N N'$ fusing into the deuteron. Since $S=1$ for the deuteron, we have three spin states and then we calculate transitions from the states $\uparrow \uparrow (1)$, $\uparrow \downarrow (2)$ to the final one $\uparrow \uparrow (1)$,
$\frac{1}{\sqrt{2}} ( \uparrow \downarrow +  \downarrow \uparrow ) (2),$ $\downarrow \downarrow (3)$.
\begin{eqnarray}
Q_{11} &:&  \uparrow \uparrow ~ \to ~ \uparrow \uparrow,
\nonumber\\
Q_{12} &:&  \uparrow \uparrow ~ \to ~ \frac{1}{\sqrt{2}} ( \uparrow \downarrow +  \downarrow \uparrow ),
\nonumber\\
Q_{13} &:&  \uparrow \uparrow ~ \to ~ \downarrow \downarrow,
\nonumber\\
Q_{21} &:&  \uparrow \downarrow ~ \to ~ \uparrow \uparrow,
\nonumber\\
Q_{22} &:&  \uparrow \downarrow ~ \to ~ \frac{1}{\sqrt{2}} ( \uparrow \downarrow +  \downarrow \uparrow ),
\nonumber\\
Q_{23} &:&  \uparrow \downarrow ~ \to ~ \downarrow \downarrow.
\end{eqnarray}
The transitions for $\downarrow \uparrow $ and $\downarrow \downarrow$ to the final states after summing over the final deuteron spins in the cross sections give the same contribution as from $\uparrow \downarrow$ and $\uparrow \uparrow$ and are accounted for multiplying the cross section by two. To evaluate the spin matrix elements, we write
\begin{eqnarray}
& &
\sigma_+ = \frac{1}{2} (\sigma_x + i \sigma_y ) ;~~~
\sigma_- = \frac{1}{2} (\sigma_x - i \sigma_y ) ;~~~
\sigma_0 = \sigma_{z}, \nonumber \\
& &
\sigma_+ | \uparrow \rangle = 0;~~~
\sigma_+ | \downarrow \rangle =  |\uparrow \rangle ;~~~
\sigma_- | \uparrow \rangle = |\downarrow \rangle ;~~~
\sigma_- | \downarrow \rangle =  0 ;~~~
\sigma_0 | \uparrow \rangle = |\uparrow \rangle ;~~~
\sigma_0 | \downarrow \rangle =  - |\downarrow \rangle, 
\nonumber\\
& &
q_+ = q_x + i q_y ;~~~ 
q_- = q_x - i q_y ;~~~ 
q_0 = q_z\ .
\label{eq:AppB_sigma_q}
\end{eqnarray}
Then,
\begin{equation}
 \vec{\sigma} \cdot \vec{q} = \sigma_+ q_- + \sigma_- q_+ + \sigma_z q_z.
\label{eq:AppB_sigmaq}
\end{equation}
Similarly we can write
\begin{equation}
 \epsilon_{ijk}\, p_{\pi i} \, q_{j} \, \sigma_k 
= \sigma_+ a_- + \sigma_- a_+ + \sigma_z a_z,\label{eq:onesig}
\end{equation}
with 
\begin{eqnarray}
a_+ &=& p_{\pi y}\, q_{z} - q_{y} \, p_{\pi z}  - i \, (p_{\pi x} \, q_{z} - q_{x} \, p_{\pi z}  ),  \nonumber\\
a_- &=& p_{\pi y}\, q_{z} - q_{y} \, p_{\pi z}  + i \, (p_{\pi x} \, q_{z} - q_{x} \, p_{\pi z}  ),  \nonumber\\
a_z &=& p_{\pi x}\, q_{y} - q_{x} \, p_{\pi y}.
\label{eq:AppB_apmz}
\end{eqnarray}
We take the direction of the incoming proton with momentum $\vec{p}$ as the $z$ direction. Without loss of generality, we can choose the $x$ and $y$ axes such that $\phi_\pi =0$. Then,
\begin{eqnarray}
 \vec{p} =  p  \left(
    \begin{array}{c}
      0  \\
      0  \\
      1 
    \end{array}
  \right);~~~
 \vec{p}_\pi =  p_\pi  \left(
    \begin{array}{c}
      \sin \theta_\pi  \\
      0  \\
      \cos \theta_\pi 
    \end{array}
  \right);~~~
 \vec{q} =  q  \left(
    \begin{array}{c}
      \sin \theta \, \cos \phi  \\
      \sin \theta \, \sin \phi  \\
      \cos \theta
    \end{array}
  \right),
\label{eq:AppB_def_moment}
\end{eqnarray}
and 
\begin{equation}
 \int d^3 q = \int q^2 dq \int^{\pi}_{0} \sin \theta d \theta  \int^{2\pi}_{0} d \phi.
\end{equation}

\subsection{Pion exchange}\label{App:B.1}
We use the property
\begin{equation}
 \sum_{M_\Delta} S_i | M_\Delta \rangle  \langle  M_\Delta | S^{\dag}_j  
= \frac{2}{3} \delta_{ij} - \frac{i}{3} \epsilon_{ijk}\, \sigma_k,
\end{equation}
and we then find, using Eqs. (\ref{eq:AppB_sigmaq}) and (\ref{eq:onesig}), for the $\Delta$-up and $\Delta$-down terms of Eq.~(\ref{eq:t_Delta}), $\vec{S}_1 \cdot \vec{p}_{\pi} \ \vec{S}_1^{\dag} \cdot \vec{q} \ \vec{\sigma}_2 \cdot \vec{q} $ and $ \vec{\sigma}_1 \cdot \vec{q} \ \vec{S}_2 \cdot \vec{p}_{\pi} \ \vec{S}_2^{\dag} \cdot \vec{q} $, the following expressions
\begin{eqnarray}
Q_{11}^{\rm up} &=& \left( \frac{2}{3} \vec{p}_\pi \cdot \vec{q} - \frac{i}{3} a_z \right)\, q_z,
\nonumber\\
Q_{12}^{\rm up} &=& \frac{1}{\sqrt{2}} \left\{ \left( \frac{2}{3} \vec{p}_\pi \cdot \vec{q} - \frac{i}{3} a_z \right)\, q_+ - \frac{i}{3} a_+ q_z \right\},
\nonumber\\
Q_{13}^{\rm up} &=&  - \frac{i}{3}\, a_+ \, q_+,
\nonumber\\
Q_{21}^{\rm up} &=& \left( \frac{2}{3} \vec{p}_\pi \cdot \vec{q} - \frac{i}{3} a_z \right)\, q_-,
\nonumber\\
Q_{22}^{\rm up} &=& \frac{1}{\sqrt{2}} \left\{ -\left( \frac{2}{3} \vec{p}_\pi \cdot \vec{q} - \frac{i}{3} a_z \right)\, q_z - \frac{i}{3} a_+ q_- \right\},
\nonumber\\
Q_{23}^{\rm up} &=&   \frac{i}{3}\, a_+ \, q_z,
\nonumber\\
Q_{11}^{\rm down} &=& \left( \frac{2}{3} \vec{p}_\pi \cdot \vec{q} - \frac{i}{3} a_z \right)\, q_z,
\nonumber\\
Q_{12}^{\rm down} &=& \frac{1}{\sqrt{2}} \left\{ \left( \frac{2}{3} \vec{p}_\pi \cdot \vec{q} - \frac{i}{3} a_z \right)\, q_+ - \frac{i}{3} a_+ q_z \right\},
\nonumber\\
Q_{13}^{\rm down} &=&  - \frac{i}{3}\, a_+ \, q_+,
\nonumber\\
Q_{21}^{\rm down} &=& -\frac{i}{3}\, a_- \, q_z,
\nonumber\\
Q_{22}^{\rm down} &=& \frac{1}{\sqrt{2}} \left\{ \left( \frac{2}{3} \vec{p}_\pi \cdot \vec{q} + \frac{i}{3} a_z \right)\, q_z - \frac{i}{3} a_- q_+ \right\},
\nonumber\\
Q_{23}^{\rm down} &=&   \left( \frac{2}{3} \vec{p}_\pi \cdot \vec{q} + \frac{i}{3} a_z \right)\, q_+.
\end{eqnarray}
 There is one more thing to be done. The $\vec{S} \cdot \vec{p}_{\pi} \ \vec{S}^{\dag} \cdot \vec{q} $ operators should be evaluated in the $\Delta$ rest frame. For this, we need to make a boost of the variables $\vec{p}_\pi$, $\vec{q}$ to the $\Delta$ rest frame. The general boost is shown in Appendix~\ref{App:C}.
As seen in Appendix~\ref{App:C}, the momenta $\vec{p}_\pi$, $\vec{q}$ which come from the operators $\vec{S} \cdot \vec{p}_{\pi} \ \vec{S}^{\dag} \cdot \vec{q} $  have to be boosted to the $\Delta$ rest frame and become $\vec{p}\,'_\pi$, $\vec{q}\,'$. This means that in the functions $Q_{ij}^{\rm up}$, $Q_{ij}^{\rm down}$ the term $\vec{p}_\pi \cdot \vec{q}$ becomes $\vec{p}\,'_\pi \cdot \vec{q}\,'$ and $p_\pi$ and $q$ in the definition of $a_+$, $a_-$, $a_z$ in Eq.~(\ref{eq:AppB_apmz}), $\vec{p}_\pi$ and $\vec{q}$ must be substituted by $\vec{p}\,'_\pi$, $\vec{q}\,'$.

In order to evaluate the transition from $\frac{1}{\sqrt{2}}(\uparrow\downarrow\pm \downarrow \uparrow)$ to $\frac{1}{\sqrt{2}}(\uparrow\downarrow+ \downarrow \uparrow)$ we need the transition from $\downarrow\uparrow$ to $\frac{1}{\sqrt{2}}(\uparrow\downarrow+\downarrow\uparrow)$, which is given by
\begin{equation}
 Q^\mathrm{up}_{32}=\frac{1}{\sqrt{2}}\lbrace(\frac{2}{3}\vec{p}_\pi\vec{q}+\frac{i}{3}a_z)q_z-\frac{i}{3}a_-q_+\rbrace\ .
\end{equation}
 Similarly, for the $\Delta$-down mechanism we have
\begin{equation}
 Q^\mathrm{down}_{32}=\frac{1}{\sqrt{2}}\lbrace(-\frac{2}{3}\vec{p}_\pi\vec{q}+\frac{i}{3}a_z)q_z-\frac{i}{3}a_+q_-\rbrace\ ,
\end{equation}
with the same boost for $\vec{p}_\pi$ and $\vec{q}$ as before.
The matrix elements for $\frac{1}{\sqrt{2}}(\uparrow\downarrow\pm\downarrow\uparrow)\to\frac{1}{\sqrt{2}}(\uparrow\downarrow+\downarrow\uparrow)$ are given by
\begin{eqnarray}
&& Q^\mathrm{up}_{\pm}=\frac{1}{\sqrt{2}}(Q^\mathrm{up}_{22}\pm Q^\mathrm{up}_{32})\nonumber\\
 && Q^\mathrm{down}_{\pm}=\frac{1}{\sqrt{2}}(Q^\mathrm{down}_{22}\pm Q^\mathrm{down}_{32})\nonumber\\
\end{eqnarray}

\subsection{Correlations term, $g'$}\label{App:B.2}
We must evaluate the matrix elements of $\vec{S}_1\cdot\vec{p}_\pi S^\dagger_1\cdot\vec{\sigma}_2$ for the $\Delta$-up mechanism and $\vec{S}_2\cdot\vec{p}_\pi S^\dagger_2\cdot\vec{\sigma}_1$ for the $\Delta$-down mechanism. We have now
\begin{eqnarray}
 \vec{S}_1\cdot\vec{p}_\pi S^\dagger_1\vec{\sigma}_2=(\frac{2}{3}\delta_{ij}-\frac{i}{3}\epsilon_{ijk}\sigma_{1,k})p_{\pi, i}\sigma_{2,j}
\end{eqnarray}
and $\epsilon_{ijk}\sigma_{1,k}p_{\pi,i}\sigma_{2,j}$ can be written in terms of $\sigma_+$, $\sigma_-$, $\sigma_z$ as 
\begin{eqnarray}
 \epsilon_{ijk} p_{\pi,i}\sigma_{2,j}\sigma_{1,k}=&&-2ip_{\pi,z}\sigma_{1,+}\sigma_{2,-}+ip_{\pi,-}\sigma_{1,+}\sigma_{2,z}\nonumber\\
 &&+2ip_{\pi,z}\sigma_{1,-}\sigma_{2,+}-i p_{\pi,+}\sigma_{1,-}\sigma_{2,z}\nonumber\\
 &&-ip_{\pi,-}\sigma_{1,z}\sigma_{2,+}+ip_{\pi,+}\sigma_{1,z}\sigma_{2,-}\label{eq:twosig}
\end{eqnarray}
by means of which we easily find
\begin{eqnarray}
 &&Q^\mathrm{'(up)}_{11}=\frac{2}{3}p'_{\pi,z};\quad Q^\mathrm{'(up)}_{12}=\frac{2}{3\sqrt{2}}p'_{\pi,+}\nonumber\\
 &&Q^\mathrm{'(up)}_{13}=0;\quad
 Q^\mathrm{'(up)}_{21}=\frac{1}{3}p_{\pi,-}'\nonumber\\
&& Q^\mathrm{'(up)}_{22}=0;\quad Q^\mathrm{'(up)}_{23}=\frac{1}{3}p'_{\pi,+}
\label{eq:qs}
\end{eqnarray}
and for the $\Delta$-down mechanism we find
\begin{eqnarray}
 Q^\mathrm{'(down)}_{ij}=Q^\mathrm{'(up)}_{ij}\ ,
\end{eqnarray}
and, as before, $\vec{p}_\pi$ has been boosted to $\vec{p}\,'_\pi$ in the $\Delta$ rest frame.
\subsection{$\rho$-exchange}\label{App:B.3}
In Eq. (\ref{eq:dups}) we found that the spin operator for $\rho$-exchange is $\vec{S}_1\cdot\vec{p}\,'_\pi(S^\dagger_1\times\vec{q}\,')\cdot (\vec{\sigma}_2\times\vec{q})$ for the $\Delta$-up mechanism and the same for $\Delta$-down exchanging $1$ by $2$. For $\Delta$-up we have,
\begin{eqnarray}
 \vec{S}_1\cdot\vec{p}\,'_\pi(S_1^\dagger\times \vec{q}\,')(\vec{\sigma}_2\times \vec{q})=&&\frac{2}{3}\vec{q}\,'\cdot\vec{q}\,\vec{p}\,'_\pi\cdot\vec{\sigma}_2-\frac{2}{3}\vec{p}\,'_\pi\cdot\vec{q}\,\vec{\sigma}_2\cdot \vec{q}\,'\nonumber\\&&-\frac{i}{3}\vec{q}\,'\cdot\vec{q}\,\epsilon_{iks}\vec{p}\,'_{\pi,i}\sigma_{2,k}\sigma_{1,s}\nonumber\\&&+\frac{i}{3}\epsilon_{iks}p'_{\pi,i}q_k\sigma_{1,s}\vec{\sigma}\,_2\cdot\vec{q}\,'
\end{eqnarray}
Using the results obtained in the two former subsections we obtain:
\begin{eqnarray}
 Q_{11}^{(\rho,\mathrm{up})}=&&\frac{2}{3}\vec{q}\,'\cdot\vec{q}\,p'_{\pi,z}-\frac{2}{3}\vec{p}\,'_{\pi}\cdot\vec{q}\,q'_z+\frac{i}{3}a_zq'_z\nonumber\\
 Q_{12}^{(\rho,\mathrm{up})}=&&\frac{1}{\sqrt{2}}\lbrace\frac{2}{3}\vec{q}\,'\cdot\vec{q}\,p'_{\pi,+}-\frac{2}{3}\vec{p}\,'_{\pi}\cdot\vec{q}\,q'_++\frac{i}{3}a_zq'_++\frac{i}{3}a_+q'_z\rbrace\nonumber\\
 Q_{13}^{(\rho,\mathrm{up})}=&&\frac{i}{3}a_+q'_+\nonumber\\
 Q_{21}^{(\rho,\mathrm{up})}=&&\frac{1}{3}\vec{q}\,'\cdot\vec{q}\,p'_{\pi,-}-\frac{2}{3}\vec{p}\,'_{\pi}\cdot\vec{q}\,q'_-+\frac{i}{3}a_zq'_-
\nonumber\\Q_{22}^{(\rho,\mathrm{up})}=&&\frac{1}{\sqrt{2}}\lbrace\frac{2}{3}\vec{p}\,'_\pi\cdot\vec{q}q'_{z}-\frac{i}{3}a_z q'_z+\frac{i}{3}a_+q'_-\rbrace\nonumber\\Q_{23}^{(\rho,\mathrm{up})}=&&\frac{1}{3}\vec{q}\,'\cdot\vec{q}\,p'_{\pi,+}-\frac{i}{3}a_+ q'_z\nonumber\\
\label{eq:rho}\end{eqnarray}
with $a_{+}$, $a_-$, $a_z$ given in Eq. (\ref{eq:AppB_apmz}) but $\vec{p}_\pi\to\vec{p}\,'_\pi$ in these expressions since it is boosted (the boosted $q'$ appears explicitly in the expressions). 
 Similarly for $\Delta$-down we will have
 \begin{eqnarray}
 Q_{11}^{(\rho,\mathrm{down})}=&&\frac{2}{3}\vec{q}\,'\cdot\vec{q}\,p'_{\pi,z}-\frac{2}{3}\vec{p}\,'_{\pi}\cdot\vec{q}\,q'_z+\frac{i}{3}a_zq'_z\nonumber\\
 Q_{12}^{(\rho,\mathrm{down})}=&&\frac{1}{\sqrt{2}}\lbrace\frac{2}{3}\vec{q}\,'\cdot\vec{q}\,p'_{\pi,+}-\frac{2}{3}\vec{p}\,'_{\pi}\cdot\vec{q}\,q'_++\frac{i}{3}a_zq'_++\frac{i}{3}a_+q'_z\rbrace\nonumber\\
 Q_{13}^{(\rho,\mathrm{down})}=&&\frac{i}{3}a_+q'_+\nonumber\\
 Q_{21}^{(\rho,\mathrm{down})}=&&\frac{1}{3}\vec{q}\,'\cdot\vec{q}\,p'_{\pi,-}+\frac{i}{3}\vec{q}\,'_{z}a_-
\nonumber\\Q_{22}^{(\rho,\mathrm{down})}=&&\frac{1}{\sqrt{2}}\lbrace-\frac{2}{3}\vec{p}\,'_\pi\cdot\vec{q}q'_{z}-\frac{i}{3}a_z q'_z+\frac{i}{3}a_-q'_+\rbrace\nonumber\\Q_{23}^{(\rho,\mathrm{down})}=&&\frac{1}{3}\vec{q}\,'\cdot\vec{q}\,p'_{\pi,+}-\frac{2}{3}\vec{p}\,'_\pi\cdot\vec{q}q'_+-\frac{i}{3}q'_+a_z\nonumber\\\end{eqnarray}
where again in $a_+$, $a_-$, $a_z$, $\vec{p}_\pi$ should be $\vec{p}\,'_\pi$ boosted according to appendix C for the $\Delta$-down mechanism.

As done for the pion exchange, we define now $\downarrow\uparrow\to\frac{1}{\sqrt{2}}(\uparrow\downarrow+\downarrow\uparrow)$ as 
\begin{eqnarray}
 &&Q^{(\rho,\mathrm{up})}_{32}=\frac{1}{\sqrt{2}}\lbrace -(\frac{2}{3}\vec{p}\,'_\pi\vec{q}+\frac{i}{3}a_z)q_z'+\frac{i}{3}a_-q_+'\rbrace\ ,\nonumber\\
 &&Q^{(\rho,\mathrm{down})}_{32}=\frac{1}{\sqrt{2}}\lbrace (\frac{2}{3}\vec{p}\,'_\pi\vec{q}-\frac{i}{3}a_z)q_z'+\frac{i}{3}a_+q_-'\rbrace\ ,
\end{eqnarray}
and the transitions from $\frac{1}{\sqrt{2}}(\uparrow\downarrow\pm \downarrow\uparrow)\to\frac{1}{\sqrt{2}}(\uparrow\downarrow+\downarrow\uparrow)$, are given by,
\begin{eqnarray}
&&Q^{(\rho,\mathrm{up})}_{\pm}=\frac{1}{\sqrt{2}}(Q^{\rho,\mathrm{up}}_{22}\pm Q^{\rho,\mathrm{up}}_{32})\nonumber\\
&&Q^{(\rho,\mathrm{down})}_{\pm}=\frac{1}{\sqrt{2}}(Q^{\rho,\mathrm{down}}_{22}\pm Q^{\rho,\mathrm{down}}_{32})
\end{eqnarray}

\section{Boost of the momenta to the $\Delta$ rest frame }\label{App:C}
In the $\pi N \to \Delta \to \pi N$ amplitude the operator $\vec{S} \cdot \vec{p}_{\pi} \ \vec{S}^{\dag} \cdot \vec{q} $ has to be evaluated in the $\Delta$ rest frame where the vertex $\delta H_{\pi N \Delta}$ of Eq.~(\ref{eq:H_piNDelta}) holds. By boosting the longitudinal component of a momentum $(m^0, \vec{m})$ from a frame where the $\Delta$ has a momentum $(E_\Delta, \vec{p}_\Delta)$ to the frame where the $\Delta$ is at rest, we obtain the formula
\begin{equation}
 \vec{m}\,' = \left[  \left( \frac{E_\Delta}{M_{\rm inv}(\Delta)} -1 \right)
\frac{ \vec{m} \cdot \vec{p}_\Delta}{|\vec{p}_\Delta|^2} - \frac{m^0}{M_{\rm inv}(\Delta)}
\right] \vec{p}_\Delta + \vec{m}  ,
\end{equation}
which we apply to $\vec{p}_\pi$ and $\vec{q}$ in this operator. In the $\Delta$-up mechanism we have, $\vec{p}_\pi = (E_\pi, \vec{p}_\pi)$, $q = (p^0 - E_\Delta, \vec{q}\, )$, ($p^0 = \frac{\sqrt{s}}{2}$),
\begin{equation}
\vec{p}_\Delta = \vec{p} - \vec{q}\, ;~~~ E_\Delta = \sqrt{s} -E_N(-\vec{p}+ \vec{q}\, );~~~ 
M^2_{\rm inv}(\Delta) = E_\Delta^2 -\vec{p}\,^2_\Delta =  s + M_N^2 - 2 \sqrt{s} E_N (-\vec{p} + \vec{q}\, ). 
\label{eq:C2}
\end{equation}
For the $\Delta$-down mechanism, we have
\begin{equation}
 \vec{p}_\Delta = -\vec{p} - \vec{q}\, ;~~~ E_\Delta = \sqrt{s} -E_N(\vec{p}+ \vec{q}\, );~~~ 
M^2_{\rm inv}(\Delta) =  s + M_N^2 - 2 \sqrt{s} E_N (\vec{p} + \vec{q}\, ).
\label{eq:C3}
\end{equation}
By performing this boost, we obtain $\vec{p}\,'_\pi$, $\vec{q}\,'$ for the $\Delta$-up and $\Delta$-down mechanism.

\section{Energy dependent $\Delta$ width}\label{App:D}
Since the $\Delta$ appears inside a loop we take the $\Delta$ width energy dependence as
\begin{equation}
 \Gamma(M_\mathrm{inv})=\Gamma_{\mathrm{on}}\frac{M_\Delta}{M_\mathrm{inv}}\left(\frac{\tilde{p}_\pi}{\tilde{p}_{\pi,\mathrm{on}}}\right)^3
\end{equation}
where 
\begin{equation}
\tilde{p}_\pi=\frac{\lambda^{1/2}(M^2_\mathrm{inv},m_\pi^2,m_\pi^2)}{2M_\mathrm{inv}}\theta(M_\mathrm{inv}-M_N-m_\pi)\,
\end{equation}
and
\begin{equation}                                                                                
\tilde{p}_{\pi,\mathrm{on}}=\frac{\lambda^{1/2}(M^2_\Delta,m^2_\pi,m^2_N)}{2M_\Delta}   \ , 
\end{equation}
with $\Gamma_\mathrm{on}$ the on-shell $\Delta$ width and $M_\mathrm{inv}$  the invariant mass of the $\Delta$, $M_\mathrm{inv}\equiv M_\mathrm{inv}(\Delta)$, which we have shown in Appendix~\ref{App:C}, Eqs.~(\ref{eq:C2}), (~\ref{eq:C3}), for the $\Delta$-up and $\Delta$-down mechanisms.

\section{The amplitudes $t_{ij}^I$ for impulse approximation}\label{App:E}
The $t_{ij}^I$ amplitudes of the  impulse approximation entering Eq.~(\ref{eq:t2}) are given explicitly by 
\begin{eqnarray}
 &&-it_{11}^I=-F_I\,p_{\pi,z}[\psi(|\vec{p}-\frac{\vec{p}_\pi}{2}|)-\psi(|-\vec{p}-\frac{\vec{p}_\pi}{2}|)]\nonumber\\
 &&-it_{12}^I=-F_I\,\frac{1}{\sqrt{2}}p_{\pi,+}[\psi(|\vec{p}-\frac{\vec{p}_\pi}{2}|)-\psi(|-\vec{p}-\frac{\vec{p}_\pi}{2}|)]\nonumber\\
 &&-it_{13}^I=0\nonumber\\
 &&-it_{21}^I=F_I\,p_{\pi,-}\psi(|-\vec{p}-\frac{\vec{p}_\pi}{2}|)\nonumber\\
 &&-it_{22}^I=-F_I\,\frac{1}{\sqrt{2}}p_{\pi,z}[\psi(|\vec{p}-\frac{\vec{p}_\pi}{2}|)+\psi(|-\vec{p}-\frac{\vec{p}_\pi}{2}|)]\nonumber\\
 &&-it_{23}^I=-F_I\,p_{\pi,+}\psi(|\vec{p}-\frac{\vec{p}_\pi}{2}|)\ ,
\end{eqnarray}
with $p_{\pi,+}=p_{\pi,x}+ip_{\pi,y }$ and $p_{\pi,-}=p_{\pi,x}-ip_{\pi,y }$.

\section*{ACKNOWLEDGMENT}
We would like to thank Igor I. Strakovsky for useful discussions and information. The work of N. I. was partly supported by JSPS Overseas Research Fellowships and JSPS KAKENHI Grant Number JP19K14709. R. M. acknowledges support from the CIDEGENT program with Ref. CIDEGENT/2019/015 and from the spanish national grant PID2019-106080GB-C21. 
This work is partly supported by the Spanish Ministerio de Economia y Competitividad and European FEDER funds under Contracts No. FIS2017-84038-C2-1-P B and No. FIS2017-84038-C2-2-P B. This project has received funding from the European Unions Horizon 2020 research and innovation programe under grant agreement No 824093 for the **STRONG-2020 project.

\end{document}